\documentclass[]{aa}
\usepackage{graphicx}
\bibpunct{(}{)}{;}{a}{}{,} % to follow the A&A style
\usepackage[utf8]{inputenc}
\usepackage{gensymb}
\usepackage{subcaption}
\usepackage{comment}
\usepackage{float}
\usepackage{color}
\newfont{\nlx}{cmssdc10 scaled 900}
\newfont{\mlx}{cmssdc10 scaled 1032}
\definecolor{myred1}{rgb}{0.,0.,0.}
\newfont{\mfont}{cmssdc10 scaled 760}
\newcommand{\com}[1]{\textcolor{myred1}{\rm #1}}
\def\ha{H$\alpha$}
\def\ewha{EW(H$\alpha$)}
\def\n2ha{log[N{\sc ii}]/H$\alpha$}

\newcommand{\kthree}{\ensuremath{k_{3,5}/k_{1}}}
\newcommand{\diffpa}{\ensuremath{\Delta \phi}}
\newcommand{\stdevpa}{\ensuremath{\sigma _{\textrm{PA}}}}
\newcommand\T{\rule{0pt}{2.6ex}}       % Top strut
\newcommand\B{\rule[-1.2ex]{0pt}{0pt}} % Bottom strut

\title{Ionized gas kinematics of massive elliptical galaxies in CALIFA and in cosmological zoom-in simulations}
\titlerunning{Ionized gas kinematics of massive elliptical galaxies}
\author{ 
Jan Florian\inst{1}
\and Bodo Ziegler\inst{1}
\and Michaela Hirschmann\inst{2,1}
\and Polychronis Papaderos\inst{3,4,5,1}
\and Ena Choi\inst{6}
\and Matteo Frigo\inst{7}
\and Jean-Michel Gomes\inst{3}
\and Rachel S. Somerville\inst{8}
} 
\authorrunning{Florian et al.}

\institute{University of Vienna, Department of Astrophysics, T\"urkenschanzstrasse 17, A-1180 Vienna, Austria
\and{DARK, Niels Bohr Institute, University of Copenhagen, Lyngbyvej 2, DK-2100 Copenhagen, Denmark}
\and Instituto de Astrof{\'\i}sica e Ci\^encias do Espa\c{c}o, Universidade do Porto, CAUP, Rua das Estrelas, 4150-762 Porto, Portugal
\and Instituto de Astrof{\'\i}sica e Ci\^encias do Espa\c{c}o, Universidade de Lisboa, OAL, Tapada da Ajuda, PT1349-018 Lisbon, Portugal
\and Departamento de F{\'\i}sica, Faculdade de Ci\^encias, Universidade de Lisboa, Edif{\'\i}cio C8, Campo Grande, PT1749-016 Lisbon, Portugal
\and Department of Astronomy, Columbia University, New York, NY 10027, USA
\and Max-Planck-Institut fuer Astrophysik, Karl-Schwarzschild-Strasse 1, 85741 Garching, Germany
\and Center for Computational Astrophysics, Flatiron Institute, 162 5th Ave, New York, NY 10010, USA
}

\date{Nov 2019}

\abstract{
Powerful Active Galactic Nuclei (AGN) are supposed to play a key regulatory role on the evolution of their host galaxies 
by shaping the thermodynamic properties of their gas component. However, little is known as to the nature and the visibility timescale of the 
kinematical imprints of AGN-driven feedback. 
Gaining theoretical and observational insights into this subject is indispensable for a thorough understanding 
of the AGN-galaxy co-evolution and could yield empirical diagnostics for the identification of galaxies that have experienced a major AGN episode in the past.   

}{% aims
We present an investigation of kinematical imprints of AGN feedback on the Warm Ionized gas Medium (WIM) of  
massive early-type galaxies (ETGs). To this end, we take a two-fold approach that involves a comparative analysis of 
\ha\ velocity fields in 123 local ETGs from the \textsf{CALIFA} integral field spectroscopy survey with 20 simulated galaxies from high-resolution hydrodynamic
cosmological \textsf{SPHgal} simulations. 
The latter were re-simulated for two modeling setups, one with and another without  AGN feedback.
}{% methods
In order to quantify the effects of AGN feedback on gas kinematics we measure three parameters that probe deviations from simple regular rotation using the \textsf{kinemetry} package. 
These indicators trace the possible presence of distinct kinematic components in Fourier space (\kthree),
variations in the radial profile of the kinematic major axis (\stdevpa),
and offsets between the stellar and gas velocity fields (\diffpa). 
These quantities are monitored in the simulations from a redshift 3 to 0.2 to assess the connection between black hole accretion history, stellar mass growth and kinematical perturbation of the WIM.
 }{% results
Observed local massive galaxies show a broad range of irregularities, 
indicating disturbed warm gas motions, irrespective of being classified via diagnostic lines as AGN or not.
Simulations of massive galaxies with AGN feedback generally exhibit higher irregularity parameters than without AGN feedback, more consistent with observations. 
Besides AGN feedback, other processes like major merger events or infalling gas clouds can lead to elevated irregularity parameters, but they are typically of shorter duration. 
More specifically, \kthree\
is most sensitive to AGN feedback, whereas 
\diffpa\ is most strongly affected by gas infall.
}{% conclusion (optional)
We conclude that even if the general disturbance of the WIM velocity is not a unique indicator for AGN feedback, irregularity parameters, high enough to be consistent with observations, can only be reproduced in simulations with AGN feedback. Specifically, an elevated value for the deviation from simple ordered motion is a strong sign for previous events of AGN activity and feedback.
}


\begin{document}

\maketitle

\section{Introduction}\label{introduction}

Soon after the discovery that most if not all massive galaxies host supermassive black holes (SMBHs)
in their centers, 
BH masses were also found to be correlated with properties of their host galaxies (such as the bulge mass and the stellar velocity dispersion, e.g.
\citealt{silk98,gebhardt00,merritt2000}).
\textrm{
Such scaling relations point to co-evolution between SMBHs and their host
galaxies with 
}
a considerable mutual influence 
(see the reviews by \citealt{kormendy2013} and \citealt{heckman2014}). 
\textrm{
This interaction can be split into processes fueling the SMBH creating
powerful Active Galactic Nuclei (AGN)
and feedback by the AGN onto the surrounding galaxy.
Part of the accreted matter is transformed into 
prodigious energy output
heating up and expelling cold gas from galaxy cores, 
and eventually preventing its re-accretion and, consequently, shutting down any star formation (SF)
\citep{bower17,raouf19}.
}

\textrm{
In fact, most cosmological simulations nowadays implement AGN feedback in some way in order to achieve e.g. a match between calculated
and observed luminosity/mass functions for massive galaxies, groups, and clusters
(for a review, see \citealt{sm12}).
Likewise, it is 
}
necessary for predicting realistic numbers of massive early-type galaxies (ETGs) with quenched nuclei and hot gas halos 
(e.g. \citealt{dubois2016,penoyre2017}). 

\textrm{
On the other hand, direct observational evidence for an interaction between central SMBH and host galaxy are scarce.
}
For example, X-ray imaging of massive elliptical galaxies  showed heated bubbles in the surrounding medium 
(see e.g. the review by \citealt{fabian2012}).
\textrm{
Parsec-scale resolution observations enabled by ALMA revealed molecular outflows very close to the center 
\citep{audibert19}.
}

In the optical regime some galaxies display bi-symmetric features in the maps of the equivalent widths (EW) of strong emission lines,
which are thought to be due to centrally-driven winds \citep{kehrig2012,cheung2016}. 
In fact, already earlier work  uncovered a faint warm ionized gas interstellar medium (WIM) in several local ETGs
(e.g. \citealt{demoulin-ulrich1984,phillips1986,kim1989}).
\textrm{
Later, integral-feld spectroscopy (IFS) studies such as the 
SAURON \citep{dezeeuw2002},
ATLAS$^{3D}$ \citep{cappellari11},
MaNGA \citep{bundy15},
CALIFA \citep{sanchez2012,sanchez2016},
and SAMI \citep{croom12}
surveys
showed that the WIM is ubiquitous in ETGs
indicating that not all of the gas is driven out of a galaxy despite each one having hosted an AGN at some point of its evolution.
Such IFU data can therefore be used to construct both stellar and gas velocity fields (VFs) to investigate intrinsic kinematic features
to characterize galaxies.
\citealt{sande17}, e.g., use measurements of anisotropy and higher-order kinematics to classify galaxies.
}

\textrm{
VFs have great potential, too, to hold generic information of effects from AGN feedback
since stellar orbits are collisionless \citep{binney1987} keeping memory of possible previous disturbances
while the warm gas component reacts to ongoing processes as gas clouds are collisional and dissipative 
(\citealt{falcon-barroso2006} and references therein).
In a SAURON-based study of Seyfert galaxies,
\citealt{dumas2007}, e.g., observed gas velocity fields with disturbances in the nuclear regions but not in the
stellar velocity fields.
In this study, we examine whether VFs provide diagnostics that can be applied to the general population of galaxies 
that all contain SMBHs that had various phases of AGN activity during their evolution.
In particular, we here examine  massive ETGs to look for kinematical patterns that could hold clues to past or ongoing AGN activity. 
To this purpose, we derive quantitative irregularity parameters of VFs 
that were introduced to quantify the influence of interactions on cluster star-forming disk galaxies
\citep{ZKRBK09,kutdemir2008,kutdemir2010}. }
This analysis method is explained in Sect.\,\ref{kinemetry}.

\textrm{
In order to explore these kinematic features in a systematic way 
we take a two-fold approach and analyze both observational data (Sect.\,\ref{obs}) 
and snapshots from numerical simulations (Sect.\,\ref{sims}) 
treated in the same way as the observational data cubes.
}
We drew a sample of 123 ETGs from the CALIFA survey of local galaxies that were processed with the 
pipeline {\sc Porto3D} (Sect.~\ref{obssample}).

For the model galaxies, we exploit cosmological simulations using SPHGal 
that incorporate physically motivated implementations of AGN feedback.
In previous papers it was already shown that this concept 
produces massive galaxies whose properties (e.g. sizes, SMBH and stellar masses, star formation rates, gas fractions) are in good agreement with observations 
(e.g. \citealt{hirschmann2014,sijacki2015,schaye2015,hirschmann2017,choi2017}).
We pick a sample of 20 ETGs from zoom-in simulations that were performed in two separate runs,
one without an AGN and one with AGN and their feedback implemented.
\textrm{
Exactly the same 20 model galaxies were already explored by \citealt{frigo2018} to investigate the impact of AGN
on the \textit{stellar} kinematics (orbits) and stellar populations.
}
We extract various snapshots that are post-processed to treat the WIM gas phase in a way that allows for 
direct comparison with observations.
We follow their time evolution knowing their mass assembly and SMBH accretion histories.
That way we can systematically explore the evolution of the gas velocity fields that are subject to a variety 
of growth processes in addition to effects from a central AGN.

Finally, in Sect.\,\ref{comp}, we compare the results from our observational and numerical analysis
and summarize our results and conclusions in Sect.\,\ref{conc}. 

\section{Theoretical framework and methods}\label{methods}

In this section, we describe \textsf{kinemetry}, the method we use to analyze the kinematics of the warm gas in our data sample (Subsection \ref{kinemetry}). We also discuss how we determine the level of current AGN activity in simulated and observed galaxies (Sect.\ref{whantheory}).

\subsection{Kinemetry}\label{kinemetry}

In order to study the effects of AGN feedback on gas kinematics, we require spatially resolved maps of the gas velocities. The means by which these maps are generated for our simulated and observed data sets are described in 
Sect. \ref{sims} and \ref{obs}, respectively. Once the velocity fields are available, we analyze them based on the kinemetry method introduced by \citet{krajnovic2006}, which works as follows.
The observed line-of-sight (los) velocity of a rotating disk is given by

\begin{equation}\label{losv}
    V_{\textrm{los}}(R) = V_{\textrm{sys}} + V_{\textrm{rot}}(R) \cos \psi \sin i,
\end{equation}

where $V_{\textrm{sys}}$ and $V_{\textrm{rot}}$ denote, respectively, the systemic and rotational velocity, $i$ is the inclination angle, $R$ the projected radius and $\psi$ the azimuthal angle.

It can be seen from this equation that, for a given orbit, the observed los velocity can be expressed as a simple function of the cosine of the azimuthal angle. Equation \ref{losv} can then be generalized to a Fourier series of the form

\begin{equation}
    V(R, \psi) = A_0(R) + \sum_{n=1}^N A_{n}(R) \sin(n \psi) + B_{n}(R) \cos(n \psi)
\end{equation}

For each radius of a velocity map, a grid representing each possible ellipse position angle and flattening is generated. The best-fitting ellipse is determined by minimizing the coefficients $A_1$, $A_2$, $B_2$, $A_3$, and $B_3$. 
The values of the coefficients of this best-fitting ellipse contain information regarding the fitted kinematics. 

In general, the $A_n$ and $B_n$ coefficients do not describe the same kinematic properties. For triaxial systems, such as the ETGs in our sample, however, we can combine the $A_n$ and $B_n$ coefficients of the same order, in the form

\begin{equation}
    k_{n} = \sqrt{A_n^2+B_n^2}
\end{equation}

Lower-order coefficients (i.e. $k_1 = \sqrt{A_1^2+B_1^2}$) describe the bulk motion (in this case simple rotation), while higher-order coefficients ($k_3$ and $k_5$) describe deviations from simple rotation. The values of the even-order coefficients ($k_2$ and $k_4$) should be practically zero for point-anti-symmetric velocity fields, such as ours, and so we do not fit them. In addition, the position angle and flattening of the best-fit ellipse give additional information as to e.g. the orientation and inclination of a rotating disk.

To quantify irregularities in a velocity field (VF) we exploit the analysis by kinemetry with respect to three parameters, first defined in \citet{kutdemir2008,kutdemir2010}, as follows:

\begin{itemize}
    \item \kthree\ = $\sqrt{A_3^2+B_3^2+A_5^2+B_5^2}/\sqrt{A_1^2+B_1^2}$, the degree of deviation from simple rotation, normalized to the amplitude of simple rotation of the gas kinematics, 
    \item \diffpa, the difference between the stellar and gas kinematic position angle (PA), and
    \item \stdevpa, the standard deviation of the gas kinematic position angles measured at different radii. 
\end{itemize}

Kinemetry was initially run on the stellar kinematics at steps of 0.1 $r_e$ from 0.1 to 1.5 $r_e$, where $r_e$ denotes the effective radius, 
in order to determine the stellar PAs,
then on the gas kinematics at the same radii to determine the gas PAs,
with the position angles and flattenings of the best-fitting ellipses allowed to vary between successive radii. \stdevpa\ was then calculated from the resulting list of gas 
PAs,
and a list of values for \diffpa\ was calculated by subtracting the gas position angles from the stellar position angles. Finally, we ran kinemetry on the gas kinematics again, with the position angles and flattenings of the ellipses fixed to their ``global'' (median) values, in order to calculate values for \kthree. 

We here apply the same thresholds for the parameters as given by \citet{kutdemir2008}. 
They based their definitions of ``irregularity" on a kinemetry analysis of isolated field galaxies from the Spitzer Infrared Nearby Galaxies Survey (SINGS, \citealt{daigle2006}) that did not show any signatures of past or ongoing interactions.
We thus classify a galaxy's kinematics as ``irregular'' if its maximum value for \kthree\ is equal to or above 0.15, the maximum value of \diffpa\ is equal to or above 25$\degr$, and the value of \stdevpa\ is equal to or above 20. 

The kinematic center of a map must be defined before it can be used with kinemetry. For the simulated data, this was trivial, as each simulated galaxy is positioned so that the center of the resulting map coincides with the center of the galaxy. For the observed galaxies, however, this was more difficult. There are several approaches to defining the kinematic center of a galaxy, including, e.g., finding the pixel with the steepest velocity gradient \citep{arribas1997} or the greatest velocity dispersion. Some of the harmonic coefficients ($A_0$, $A_2$, and $B_2$) are also sensitive to incorrect definitions of the center \citep{krajnovic2006}. All of these methods assume regular kinematics, however, which is not the case for all of our data. Therefore, we define the center of an observed galaxy as the point with the greatest continuum flux (within a small box around the centroid of the map, in order to avoid capturing bright, neighboring objects).

Kinemetry is programmed to automatically stop its fitting when it reaches the edge of a map. Again, this was not a problem for the simulated galaxies, which were positioned in such a way as to fully fill out their frames. For observed galaxies, it was noted that the fitting was sometimes carried out past the galaxies' edges (resulting in incorrect parameter values) if the fitting was not manually stopped beforehand. For this reason, only the pixels that fulfilled certain criteria (e.g. sufficiently accurate velocity measurements, see Section \ref{obs} for more details) were used for the fitting. For some observed galaxies, this meant that only a relatively small portion of the pixels in a map were usable, and these pixels were not necessarily distributed evenly across the map. A series of tests was carried out in order to examine the effects of excluding portions of the velocity map of a simply rotating disk on the values of the irregularity parameters. We found that the main effect of excluding portions of a map is an increase in the derived values of \kthree, whereas the position angle was not as significantly affected.

\subsection{Spectroscopic classification after WHAN}\label{whantheory}
In order to explore whether ongoing accretion-powered nuclear activity is typically accompanied by an enhanced 
level of irregular motions in the WIM, as a consequence of AGN feedback, we need to first check whether an ETG shows spectroscopic evidence for an AGN. This is usually done by means of emission-line diagnostic diagrams. 

\textcolor{black}{In this study, we use the so-called WHAN classification scheme \citep{cidfernandes2010,cidfernandes2011} to 
identify among our observed galaxies those hosting an AGN (see Sect. \ref{obssample}). 
%In a WHAN diagram, 
Galaxies are assigned to one of four categories based on their H$\alpha$ equivalent widths (\ewha) and their N{\sc ii}/H$\alpha$ flux ratios: systems with a 
log[N{\sc ii}]/H$\alpha$ $>$--0.4 are subdivided into strong AGN (\ewha$>$6 \AA), weak AGN (3 $\leq$ \ewha\ (\AA) $\la$ 6), retired (0.5 $\leq$ \ewha\ (\AA) $\la$ 3) and 
passive (\ewha$<$0.5 \AA). In our work, galaxies fulfilling the criteria of either strong or weak AGN are considered to be currently affected by the energetic output from a Seyfert nucleus.}

\textcolor{black}{We note that the nature of ETGs falling in the locus of retired or passive galaxies is a subject of controversy. 
The \ewha\ of a significant fraction of these systems 
\citep[e.g., $\sim$40\% of the sample of CALIFA ETGs studied by][]{papaderos2013} 
\citep[see also][]{Singh2013} 
lies within 1 $\leq$ \ewha\ (\AA) $\la$ 3, which is consistent with predictions from photoionization models \citep[e.g.][]{cidfernandes2011,gomes2016} for galaxies whose ionizing radiation is solely powered by an evolved ($>$100 Myr) 
post-asymptotic-giant-branch (pAGB) stellar population. This fact has been interpreted by many authors as evidence that these retired (no longer star-forming) ETGs lack even a weak AGN, as proposed by \citet{cidfernandes2010}.
However, this interpretation was disputed by \citet[][]{papaderos2013}. These authors show that, in the absence of a gaseous reservoir of sufficient density, 70-90\% of Lyman continuum (LyC) radiation produced by pAGB sources in most (60\%) CALIFA ETGs escapes into the halo without being locally reprocessed into nebular emission. 
This high LyC photon escape fraction implies therefore a radical reduction of nuclear \ewha's by $\ga$1 dex and in practice prevents detection of an AGN through optical spectroscopy. 
Additionally, \citet{papaderos2013} point out that in the triaxial geometry of ETGs, the old stellar background along the line of sight can strongly dilute nuclear EWs, thereby shifting a strong Seyfert source (\ewha\ $\gg$ 6 \AA) into the locus of retired or passive galaxies. Following the above considerations, an \ewha\ in the range 0.5--3 \AA\ is no compelling evidence against 
a strong AGN in retired galaxies (RGs) but merely a necessary (yet not sufficient) condition for pure pAGB photoionization.
For the sake of completeness, we note that ETGs falling in the class of RGs according to the WHAN classification  
are usually classifiable on Baldwin, Philips and Terlevich (BPT) diagrams \citep{bpt1981} 
as LINERs \citep[Low-Ionization Nuclear Emission Regions;][]{heckman1980b}.
}
As for our simulated galaxies, we follow two separate runs, one with BHs and corresponding feedback implemented and 
one without AGN.
For the former, we quantify the activity of the AGN by the Black Hole Accretion Rate (BHAR) in comparison to the star formation rate. The time evolution of a galaxy is characterized by the growth in stellar mass through mergers and accretion.

\section{Cosmological zoom-in simulations of massive galaxies}\label{sims}

In this section, we begin with a brief description of the hydrodynamic SPH code used to simulate our sets of massive galaxies (Sect. \ref{sphgal}), outline the simulation setup (Sect. \ref{simssetup}), and explain the generation of velocity maps of the warm ionized gas of our simulated galaxies (Sect. \ref{simsvelmaps}).

\subsection{Simulation sample}\label{simssample}

The sets of simulated galaxies, examined in this work, have been introduced by \citet{hirschmann2017} and \citet{choi2017}, and we urge readers to consult 
these works for more details beyond the brief summary we provide here.

\subsubsection{The simulation code SPHGal}\label{sphgal}

The cosmological zoom-in simulations of massive galaxies were generated using the code \textsf{SPHGal}, a modified version of the smooth particle hydrodynamics (SPH) code \textsc{GADGET3} \citep{springel2005}. As outlined in \citet{hu2014}, \textsf{SPHGal} incorporates several refinements of the classic SPH implementation, in order to pass all standard hydrodynamical tests, which have been reported to be problematic for SPH techniques in the past. 
Our simulation code also includes recipes for different baryonic processes, such as star formation, feedback from stars and AGN, chemical enrichment, metal-line cooling, as well as an ultraviolet photo-ionization background \citep{haardt2001}.

Star formation and the resulting chemical evolution (described in \citealt{aumer2013} and \citealt{nunez2017}) are modeled by a stochastic conversion of gas particles, whose densities are above a temperature-dependent threshold (thus becoming Jeans unstable) into star particles. We assume chemical enrichment via type-Ia and type-II supernovae (SNe) andAGB stars, where we trace 11 elements (H, He, C, N, O, Ne, Mg, Si, S,Ca and Fe) in both gas and star particles. For gas particles, we include metal diffusion to allow for a more realistic mixing of metals released into the ambient gas (\citealt{aumer2013}). 

Star formation is regulated by both stellar and AGN feedback. We adopt the approach outlined in \citet{nunez2017} to account for mass, energy and momentum release from young massive stars as well as AGB stars and SN (type-II and -Ia) explosions. AGN feedback is tied to the prescription for BH growth. BHs are represented by collisionless sink particles, a BH seed of $10^5 M_{\odot}$ being placed at the density minimum of any dark matter halo, whose mass exceeds $10^{11} M_\odot$. BHs can further grow via two channels: gas accretion and merger events with other BHs. Gas accretion is assumed to follow a statistical Bondi-Hoyle approach \citep{bondi1952,choi2012}.

To model AGN feedback, we rely on a physically motivated approach including both mechanical and radiative feedback \citep{ostriker2010,choi2017}. Specifically, we incorporate the effect of AGN-driven winds (motivated by observed broad-absorption-line winds) by randomly kicking gas particles in the vicinity of the BH with a velocity of 10,000 km/s perpendicular to the gaseous disk to mimic bi-conical gaseous outflows. 
\textrm{
We adopt a wind efficiency of 0.005 as described by equation 3 of Choi et al. 2017, not distinguishing between different BH accretion regimes (i.e. independent of the Eddington-ratio). 
} 
In addition, 
radiative feedback from Compton and photoionization heating due to X-ray radiation from the accreting BH, radiation pressure associated with X-ray heating and the Eddington force are also included \citep{ostriker2010}.  
In contrast to many state-of-the-art cosmological simulations, with our AGN feedback model, our simulations predict a realistic amount of hot halo gas and X-ray luminosities consistent with observations. 
\textrm{ Note that in our simulations, most of work from an accreting BHs is done in the radiatively efficient mode, i.e. when the BH is accreting at high Eddington-ratios.
}

As shown in various recent studies \citep{hirschmann2017, hirschmann2018, choi2017, choi2018, brennan2018, frigo2018}, adopting the outlined sub-grid models, and in particular the sophisticated prescription for AGN feedback, allows us to  simulate massive galaxies with many realistic properties, such as star formation histories, baryon conversion efficiencies, galaxy sizes, stellar populations, gas fractions and hot-gas X-ray luminosities.

\subsubsection{Simulation setup}\label{simssetup}

The initial conditions for the cosmological zoom-in simulations were drawn from \citet{oser2010,oser2012} and \citet{hirschmann2012,hirschmann2013}, adopting a WMAP3 cosmology (h=0.72, $\Omega _{\Lambda}$ = 0.74, $\Omega_m$ = 0.26, $\sigma_8$ = 0.77, see e.g. \citealt{spergel2003}). The dark matter (DM) haloes, chosen to be re-simulated, were selected from a "parent" dark matter-only N-body simulation, starting at z = 43, with 512$^3$ particles within a box with co-moving, periodic boundary conditions and a box length of 72 Mpc $h^{-1}$ \citep{moster2010}. At z = 0, the selected DM haloes have virial masses between 2.2 $\times$ 10$^{12}$ M$_\odot$ $h^{-1}$ and 2.2 $\times$ 10$^{13}$ M$\odot$ $h^{-1}$. All DM particles within twice the virial radius of these haloes were  traced back in time,  and then replaced with higher-resolution DM and gas particles with masses of $m_{\mathrm{dm}}$ = 2.5 $\times$ 10$^7$ M$\odot$ and $m_{\mathrm{gas}}$ = 4.2 $\times$ 10$^6$ M$\odot$, respectively \citep[see][for more details]{oser2010}.  Ten of the most massive re-simulated haloes, each of them ran with and and without AGN feedback, were finally selected to be analysed in this work.

Our AGN feedback model is very efficient in removing the majority of the warm/cool gas content of our simulated massive galaxies by $z = 0$. To be able to properly study the warm gas kinematics of galaxies, we consider the evolution of the gas content of galaxies only down to $z=0.2$ so that  a reasonably large amount of warm gas is present. A comparison to observed local massive CALIFA ETGs is then performed at $z = 0.5$. This comparison is justified, as by $z = 0.5$, our simulated galaxies have developed early-type morphologies and have masses similar to that of selected CALIFA galaxies.

\subsection{Generating mock velocity maps}\label{simsvelmaps}

In order to generate mock 2D gas velocity maps of simulated galaxies, which resemble those created via integral-field spectroscopy, we use an extended version of the publicly available PYGAD analysis package,  including a stellar population and kinematics tool as described in \citet{frigo2018}, following the analysis presented in \citet{jesseit2007,jesseit2009} and \citet{naab2014}. 

\begin{figure*} % irr params
\centering
\includegraphics[width=\linewidth]{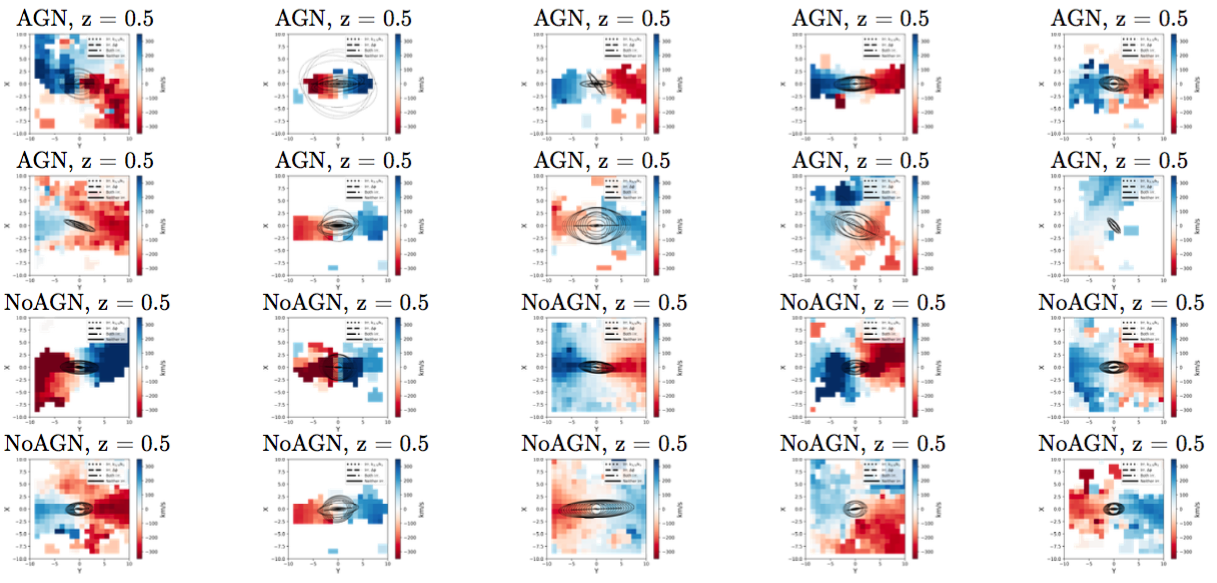}
\centering

\caption[Velocity maps of simulated galaxies at z = 0.5]{\scriptsize The velocity maps of the warm gas in the ten galaxies in our simulated data sample with (top two rows) and without (bottom two rows) AGN feedback at z = 0.5. Each galaxy is oriented edge-on, each map has an extent of 10 $\times$ 10 kpc, and the colors scale between -350 and 350 km/s. The line extending across the maps is oriented at the median position angle of the best-fitting ellipses within 1.5 r$_{e}$. The solid portion of the line has the length of the galaxy's effective radius from the center along the major axis, while the dotted portion represents the remaining fitted portion up to 1.5 r$_{e}$. The plotted ellipses are the best-fitting ellipses at every radius, with the ellipses' flattenings and position angles allowed to vary between radii. The standard deviation of these ellipses' position angles is \stdevpa, and the position angles are compared with those of the stellar velocity to compute \diffpa. \kthree\ is calculated based on a different set of ellipses at the same radii with fixed position angles (all oriented along the solid line) and flattenings (the median flattening of the first set of ellipses). The linestyles of the ellipses correspond to the values of the irregularity parameters as follows: a solid line means that both \kthree\ and \diffpa\ are within the regular ranges, a dotted line means that \kthree\ is irregular, a dashed line means that \diffpa\ is irregular, and a dashed-dotted line means that both are irregular.}
\label{fig:simsappendix2} 
\end{figure*}

With this post-processing code, positions and velocities of the simulated galaxies are centred on the densest nuclear regions using a shrinking sphere technique on the stellar component.  We  then  calculate the eigenvectors of the reduced inertia tensor (Bailin \& Steinmetz  2005)  of  all  stellar  particles  within  10~per cent  of  the virial radius, and use them to align the galaxies' principal axes with the coordinate system, such that the x-axis is the long axis and the z-axis is the short axis. 
In projection, the gas particles are mapped onto a regular two-dimensional grid, either with a pixel size of 0.4 kpc (at $z = 0$) for our examination of the galaxies' evolution histories, or with a pixel size of 1.0 kpc for comparison with the observed CALIFA galaxies to resemble more closely the average spaxel size of CALIFA IFU data.  An overview of the warm gas velocity maps generated this way is given in Fig. \ref{fig:simsappendix1} in the Appendix for the galaxies simulated with AGN feedback and in Fig. \ref{fig:simsappendix2} for those without AGN feedback. 
To finally obtain the three irregularity parameters (see sec. \ref{kinemetry}) for the warm gas component of our simulated galaxies, we apply the kinemetry tool to the warm gas velocity maps as described in section \ref{kinemetry}.

\section{Ionized gas kinematics of simulated galaxies}\label{simsresults}

In this section, we start with investigating the kinematics of the warm gas content and related properties of three example simulated galaxies, ran with and without AGN feedback. Specifically, we discuss their gas velocity maps, inclination/projection effects on their irregularity parameters as well as the cosmic evolution of corresponding galaxy stellar masses, BH accretion rates, warm gas mass fractions, and the three irregularity parameters \kthree, \diffpa, and \stdevpa. We then continue to examine the statistics of irregularity parameters for all simulated galaxies at $z=0.5$ and at $z=1.3$, ran with and without AGN feedback (Subsection \ref{simsstats}).

\subsection{Three case studies}\label{simscases}

We choose our three galaxy test cases such that their merger histories are fundamentally different: Galaxy 0094 has experienced a recent major merger at z $\sim$ 1, galaxy 0215 has had no mergers (major or minor) at all after z $\sim$ 3, and galaxy 0290 has undergone two minor mergers after z $\sim$ 2.  

\begin{figure*} % vmaps and gas maps
\centering
\includegraphics[width=\linewidth]{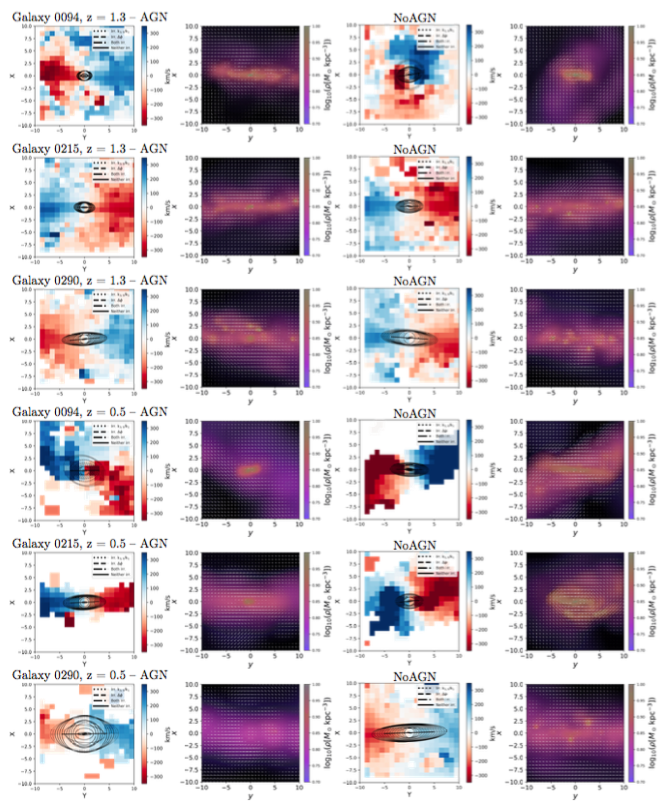}
\caption[Velocity and gas maps for halos 0094, 0215, and 0290]{\footnotesize Velocity and density maps (with velocities overplotted as white arrows) of the warm gas content of the galaxies 0094 (first and fourth row), 0215 (second and fifth row), and 0290 (third and sixth row),  at z = 1.3 (top three rows) and at z = 0.5 (bottom three rows), simulated with (first and second columns) and without  AGN feedback (third and fourth columns). The line across the velocity maps is oriented at the median position angle of the best-fitting ellipses within 1.5 r$_{e}$, the extent of the solid line illustrates the galaxy's effective radius. The plotted ellipses are the best-fitting ellipses at every radius, with the ellipses' flattenings and position angles allowed to vary between radii. Following the definitions of \com{regularity/irregularity of \citet{kutdemir2008,kutdemir2010}}, regular ranges of \kthree\ and \diffpa\ are illustrated by solid lines of the ellipses, only irregular  \kthree\ by dotted lines, only irregular \diffpa\ by dashed lines, and irregular values of both \kthree\ and \diffpa\ by dashed-dotted lines.}
\label{fig:multiplotsoverview}
\end{figure*}

In Fig. \ref{fig:multiplotsoverview}, we start with showing velocity and density maps of the warm gas of our three example galaxies 0094 (first and fourth row), 0215 (second and fifth row), and 0290 (third and sixth row) with (two left columns) and without AGN feedback (two right columns) at $z = 1.3$ (top three rows) and $z = 0.5$ (bottom three rows). In general, at $z=1.3$, warm gas motions are more irregular and regular rotation features are less strongly pronounced compared to lower redshifts such as $z=0.5$. Fig. \ref{fig:multiplotsoverview} also shows that AGN feedback has an on average more significant influence on the warm gas kinematics at $z = 0.5$ than at $z = 1.3$. Specifically, by $z=0.5$, AGN feedback prevents to a large extent the development of an extended disk-like gas structure with ordered rotation, and additionally lowers the warm gas density due to gas heating and ejection.

\begin{figure*} % 10 incs

\centering
%\text{0094}\par
\includegraphics[width=\linewidth]{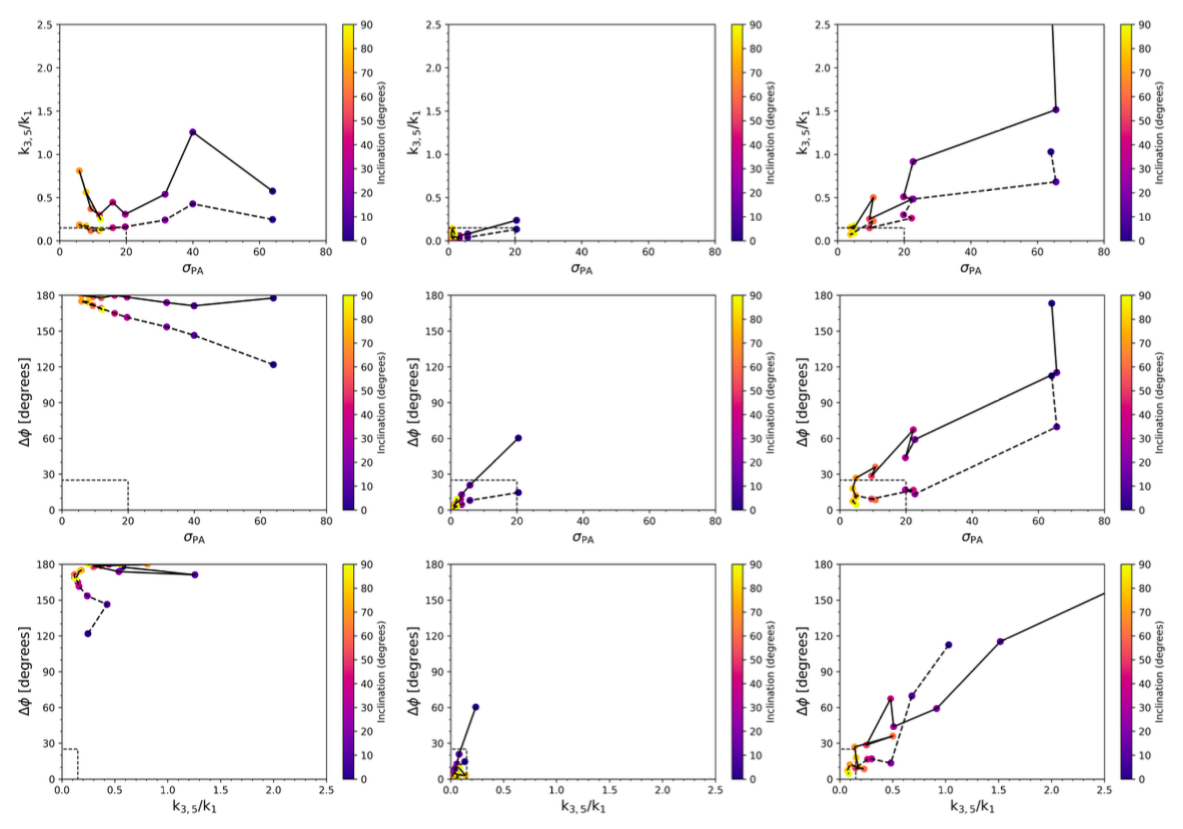}
\centering
\caption[Irregularity parameters at different inclinations]{\footnotesize Mean (dashed lines) and maximum (solid lines) \kthree\ versus \stdevpa\ (top row), \diffpa versus \stdevpa\ (middle row) and \diffpa\ versus \kthree\ (bottom row) within 1.5 r$_{e}$  of three example galaxies 0094 (left column), 0215 (middle column), and 0290 (right column), simulated with AGN feedback. In each panel, the color-coded points represent ten different inclinations of the galaxy (from face-on, 0$\degr$, to edge-on, 90$\degr$).}
\label{fig:10incs}
\end{figure*}

Taking advantage of the gas velocity maps in Fig. \ref{fig:multiplotsoverview}, we compute the mean and maximum values of the irregularity parameters \kthree, \diffpa, and \stdevpa\ within 1.5 $r_{e}$ of our three example galaxies. For these calculations, we consider ten different inclinations of the galaxies between face-on (0$\degr$) and edge-on (90$\degr$), to explicitly study the effect of inclination on the irregularity parameters. Fig. \ref{fig:10incs} depicts the mean (dashed lines) and maximum (solid lines) irregularity parameters, plotted pairwise against each other (different rows) for galaxies 0094 (left colum), 0215 (middle column), and 0290 (right column), with the inclination angle color-coded.

Fig. \ref{fig:10incs} shows that generally, the closer the galaxy's inclination is to face-on, the higher are the irregularity parameters (even if Galaxy 0094 represents an exception, as the maximum \diffpa\  stays more or less constant at $\sim180\degr$ regardless of inclination, and the mean \diffpa\ value decreases).  Such an increase of irregularity parameters from edge-on towards face-on projection implies that a galaxy could be classified in observations as regular, when viewed edge-on, while it would be deemed irregular if observed closer to face-on. This relation between irregularity parameters and inclination angle likely arises due to the fact that the velocities of the warm gas particles of these galaxies are predominantly oriented within the plane of rotation. As the observed inclination nears face-on, the component of the velocity oriented in the direction of the observer steadily decreases, largely removing any velocity gradients, which makes ordered rotation more difficult to identify. 

\begin{figure*} % multiplots
\centering
\includegraphics[width=\linewidth,height=0.525\textheight]{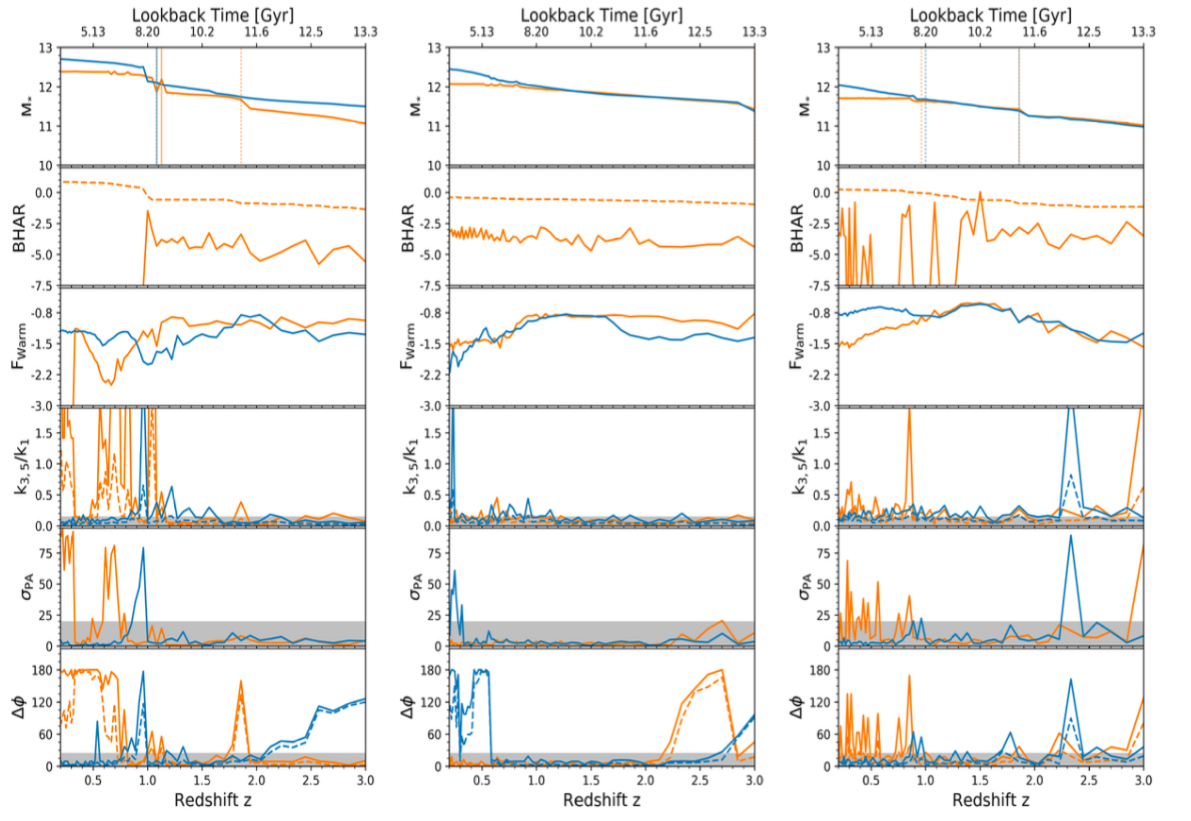}
\centering
\caption[Time evolution of halos 0094, 0215, 0290]{\footnotesize Cosmic evolution of the galaxy stellar mass within 1/10 of the virial radius, the BH accretion rate, the warm gas mass fractions as well as the irregularity parameters $k_{3,5}$/$k_1$, $\Delta \phi$, and $\sigma_{\mathrm{PA}}$ (panels from top to bottom) of our three example galaxies (0094: right column; 0215: middle column; 0290: right column) with (orange lines) and without (blue lines) AGN feedback from $z=3.0$ to $z=0.2$. In the panels showing the stellar mass histories (top row), solid lines indicate major mergers and dashed lines indicate minor mergers. The dashed lines in the panels in the second row represent the BH accretion rate at the Eddington limit. The grey areas in the panels showing the irregularity parameters (fourth to sixth rows) indicate the ``regular'' ranges (see Sec.\,2.1), and the solid (dashed) lines illustrate the maximum (mean) values of each irregularity parameter within 1.5 r$_{e}$.} \label{fig:multiplots1}
\end{figure*}

Having visually seen in Fig. \ref{fig:multiplotsoverview} that AGN feedback may affect the gas kinematics of massive galaxies towards lower redshifts, we now aim at getting a more quantitative understanding, when in cosmic history and on which time-scales AGN feedback starts to affect irregularity parameters, and to what extent this is related to host galaxy properties. Thus, in Fig. \ref{fig:multiplots1}, we show the cosmic evolution of the galaxy stellar mass, the BH accretion rate, the warm gas mass fraction, and the three irregularity parameters (panels from top to bottom) for our example galaxies 0094 (left column), 0215 (middle column) and 0290 (right column). In order to minimize the increase of the irregularity parameters caused by projection angles closer to face-on (as described above), all galaxies and their main progenitors were oriented edge-on. In the following subsections, we discuss the histories of these three example galaxies in more detail. 

%\clearpage

%\clearpage

\subsubsection{Galaxy 0094}

The galaxy 0094's stellar mass assembly history is characterized by a major merger at $z \sim 1$ (orange lines in top left panel in Fig.  \ref{fig:multiplots1}), which triggers, in the run including AGN, a strong burst of gas accretion onto the central BH (second left panel in Fig.  \ref{fig:multiplots1}). This high rate of BH accretion in turn results in a phase of significant gas heating and ejection, so that the total galaxy stellar mass remains constant after $z \sim 1$, as in-situ star formation is strongly suppressed due to the lack of available dense gas (see also \citealt{choi2017}). Also the warm gas fractions are strongly reduced after $z \sim 1$, causing subsequent fluctuations in all three irregularity parameters. 

The bottom three left panels of Fig. \ref{fig:multiplots1} show that the values of \kthree, \stdevpa, and to a lesser extent \diffpa\ are sensitive to the fraction of warm gas being present in the galaxy. When the warm gas mass fraction strongly drops between z $\sim$ 1 and z $\sim$ 0.7 (a timespan of ~1.5. Gyr), all three irregularity parameters significantly increase (into irregular ranges above the grey area). Then, as the warm gas mass fraction rises almost to its pre-feedback levels between z $\sim$ 0.7 and z $\sim$ 0.3 (a timespan of ~2.9 Gyr), \kthree\ and \stdevpa\ drop sharply into  regular ranges, while \diffpa\ remains near 180$\degr$. Finally, when the warm gas mass fraction drops to zero shortly before z = 0.3, \kthree\ and \stdevpa\ rise into the irregular ranges again. 

We explain these fluctuations in irregularity parameters as follows: AGN efficiently heats and removes warm gas (and is actually also consumed via star formation). Infalling gas clouds then replenish the warm gas content, and their angular momenta govern the further gas kinematics of the galaxy, leading to counterrotation of the gas disk with respect to the stellar motion. \kthree\ and \stdevpa\ appear to be quite sensitive to the amount of warm gas present (and thus to past AGN activity, which is largely responsible for the removal/heating of the warm gas), while \diffpa\ is likely more sensitive to changes in the angular momentum of the warm gas brought in via infalling gas clouds.

In the run with AGN feedback, the galaxy 0094 also undergoes a minor merger at z $\sim$ 2, which is associated with a spike in the irregularity parameters (in particular \diffpa), but the values of the parameters drop quickly back to nearly zero. This is likely because no significant AGN activity is triggered: the bolometric AGN luminosity at this redshift is $2.4 \times 10^{43}$~erg/s, corresponding to 0.3~ per cent of the maximum Eddington luminosity (compared to $1.9 \times 10^{45}$~erg/s, 6.8~per cent of the Eddington luminosity, at the peak in the accretion rate at $z \sim 1$).

Since AGN-driven winds originate from the BH at the center of the galaxy, we may expect that the disturbances in the gas (and thus the increased values of the irregularity parameters) should propagate radially outward over time. For Galaxy 0094, we, however, do not find a significant relation between the radius, at which \kthree\ is maximal, and the elapsed time since the peak in the BH accretion rate. This may be due to the fact that the time between output snapshots, ranging between 114 Myr and 171 Myr, is too long to properly capture the radial propagation of irregular gas motions (even if typical gas dynamical timescales are larger than the times between two snapshots, ranging between 412 Myr and 10.36 Gyr from $z=1$ to $z=0.2$).

Turning to  the run without AGN feedback, the galaxy's evolution is rather similar to that with AGN feedback until the major merger event at z $\sim$ 1. Without AGN feedback, this merger is associated with a brief spike in all three irregularity parameters, but the parameters quickly drop to values close to zero and stay low. The warm gas mass fraction is similarly unaffected overall, and the total stellar mass continues to rise as stars are able to form from available dense gas. This indicates that a merger can temporarily disturb the gas kinematics, but if there is still a large amount of gas present, disturbed gas motions quickly revert to regularity and infalling gas clouds have a much smaller effect on the kinematical behaviour (see, however, the description of Galaxy 0215 for a counterexample in the following subsection \ref{gal0215}). 

\subsubsection{Galaxy 0215}\label{gal0215}

In  both the run with and without AGN feedback, galaxy 0215 has a fairly quiet mass assembly history (top middle panel of Fig.  \ref{fig:multiplots1}),  as it does not experience any major or minor mergers after z = 2.8. In the run with AGN feedback, the BH accretion rate does not show any major peaks (second middle panel of Fig.  \ref{fig:multiplots1}), remaining mostly constant until z = 0.2, and the warm gas mass fractions do not exhibit any major decreases, typically associated with peaks in AGN activity and feedback. Consequently, the parameters stay within the regular range for the course of most of galaxy 0215's evolution (fourth, fifth and sixth middle panels of Fig.  \ref{fig:multiplots1}). 

Differences in the runs with and without AGN feedback begin to arise at z $\sim$ 0.8. While with AGN feedback, the irregularity parameters do not increase over cosmic time, in the run without AGN feedback, around $z \sim 0.6$, a gas cloud falls into galaxy 0215, as indicated by the strong increase in the warm gas mass fraction (blue line in third middle panel of Fig.  \ref{fig:multiplots1}). As a consequence, \diffpa\ in the run without AGN feedback immediately rises up to around 180$\degr$, while the other parameters stay low. Only around z $\sim$ 0.2, \kthree\ and \stdevpa\ also begin to rise, reacting to the decreasing warm gas mass fraction as a consequence of gas consumption via star formation.

We conclude that irrespective of AGN feedback, an infalling gas cloud can cause the orientation of a gas disk to change, resulting high values of \diffpa. Depending on the circumstances, the gas disk may settle in its new configuration, leading to values of \diffpa\ that remain high over time (i.e. counter-rotating gas disk), whereas the other two irregularity parameters \kthree\ and \stdevpa\  are not necessarily permanently increased without any significant continuous energy input, e.g. from an AGN. This suggests, \kthree\ and \stdevpa\ may be more sensitive to AGN feedback compared to \diffpa.

\begin{figure*}[htb!]% irr params
\centering
\includegraphics[width=\linewidth]{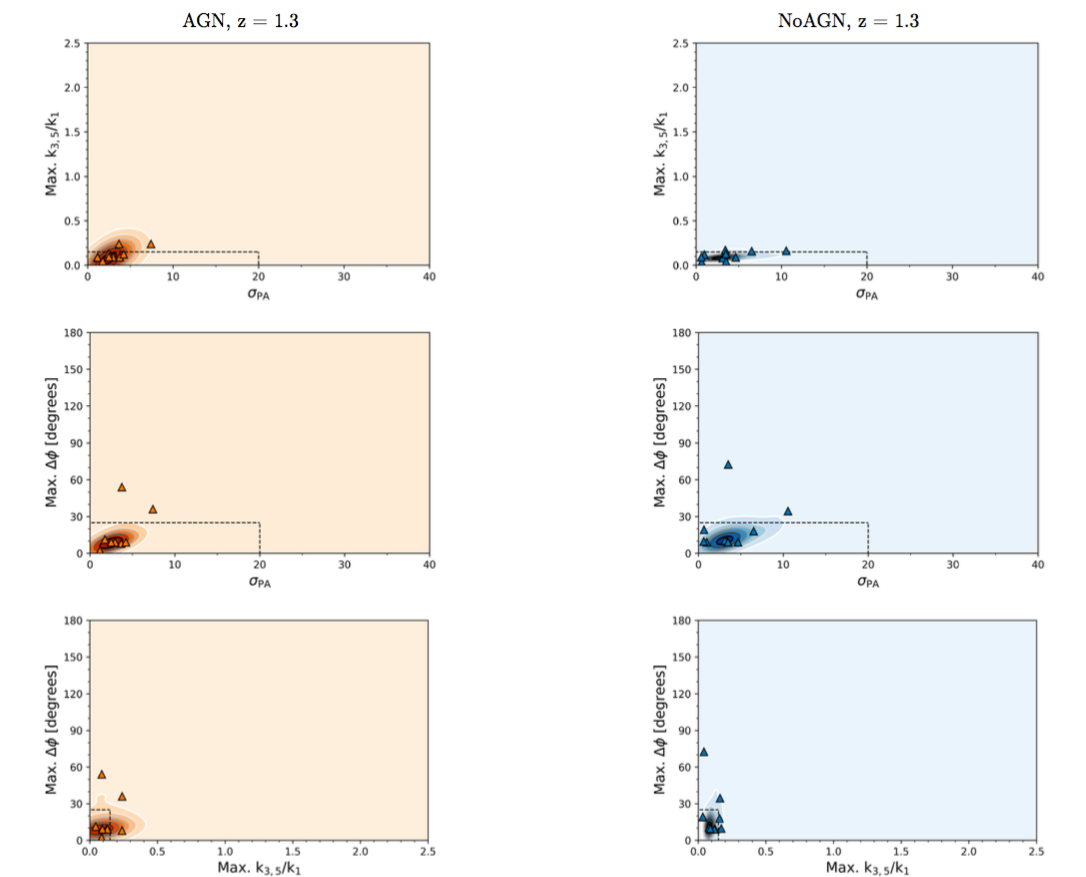}

\centering

\caption[Combinations of irregularity parameters for simulated galaxies at z = 1.3]{\footnotesize \stdevpa\ vs. \kthree\ (top row), \stdevpa\ vs. \diffpa\ (middle row), and \kthree\ vs. \diffpa\ (bottom row), for simulated galaxies with AGN feedback (left column) and without AGN feedback (right column) at z = 1.3. The triangle-shaped markers represent the maximum values within 1.5 r$_{e}$ for our galaxies inclined edge-on, while the colored contours represent the range of maximum values exhibited by each of our galaxies at 25 random orientations. The dashed lines indicate the upper boundaries of the parameter ranges within which a galaxy's kinematics are considered ``regular.''}
\label{fig:irrparamssims1} 
\end{figure*}

\begin{figure*} % irr params
\centering
\includegraphics[width=\linewidth]{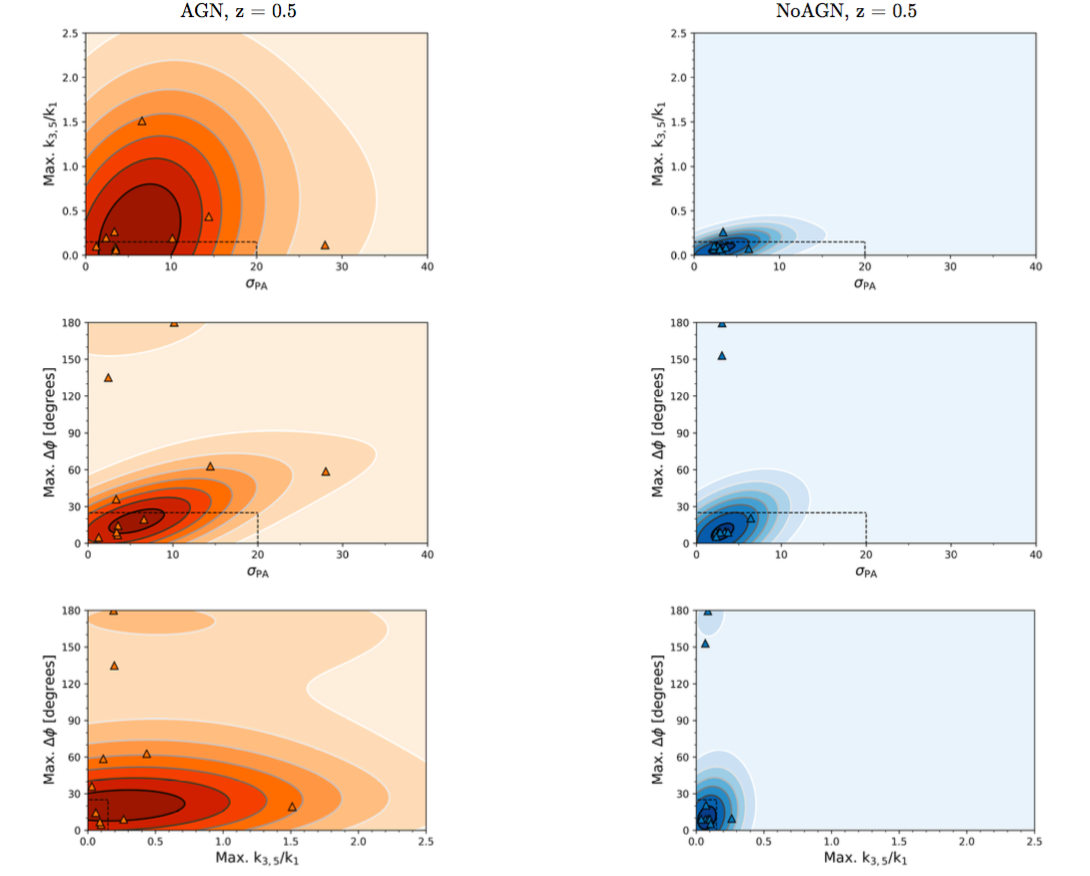}
\centering
\caption[Combinations of irregularity parameters for simulated galaxies at z = 0.5]{\footnotesize \stdevpa\ vs. \kthree\ (top row), \stdevpa\ vs. \diffpa\ (middle row), and \kthree\ vs. \diffpa\ (bottom row), for simulated galaxies with AGN feedback (left column) and without AGN feedback (right column) at z = 0.5. The triangle-shaped markers represent the maximum values within 1.5 r$_{e}$ for our galaxies inclined edge-on, while the colored contours represent the range of maximum values exhibited by each of our galaxies at 25 random orientations. The dashed lines indicate the upper boundaries of the parameter ranges within which a galaxy's kinematics are considered ``regular.''}
\label{fig:irrparamssims2} 
\end{figure*}

\subsubsection{Galaxy 0290}

The evolutionary history of galaxy 0290 is similar to that of galaxy 0094, as the former also experiences a strong burst of BH activity (temporary even with super-Eddington accretion) at $z \sim 1.5$, although in this case, the AGN peak is not associated with any major merger event (first and second right panels of   Fig.  \ref{fig:multiplots1})\footnote{This is an interesting example showing that peaks in AGN activity can be, but are not necessarily fuelled by a merger event, \citep[see][for a statistical analysis]{steinborn2018}}. 
Corresponding feedback from the accreting BH causes the warm gas mass fraction to smoothly drop after $z \sim 1.5$. 

The minor merger at $z = 1$ causes a further peak in the BH accretion rate, followed by a more severe decline in the warm gas fractions, and associated with an increase of all three irregularity parameters (fourth, fifth and sixth right panels of  Fig.  \ref{fig:multiplots1}). Even after this minor merger event, the BH of galaxy 0290 continues to episodically accrete some gas, resulting in peaks of the BH accretion rate (i.e. feedback), which we find to be clearly related with spikes in the irregularity parameters, confirming our previous conclusion that gas outflows can lead to stronger deviation from simple ordered rotation of the warm gas content. 

Without AGN feedback, the minor merger at $z \sim 1$ also causes short-lived spikes in the irregularity parameters, however, with much lower values than in the case with AGN feedback, and specifically \stdevpa\ vs. \kthree\ stay within the regular ranges (grey shaded areas and blue lines in fourth, fifth and sixth right panels of  Fig.  \ref{fig:multiplots1}). As seen before for galaxy 0094, without AGN feedback, irregularities in the gas motion are quickly reverted into ordered rotation.

From these three test cases of massive early-type galaxies, we may conclude that in general, both merger events and high levels of AGN feedback (not necessarily associated with mergers) can lead to a severe lack of coherent motions of the warm gas content, reflected by an increase of all three irregularity parameters. {\it Without} any energy/momentum release due to a central AGN, however, after a merger event, the perturbed gas motions typically quickly return to ordinary and regular rotation of the warm gas, leading to an immediate decrease of the irregularity parameters. Solely \diffpa, i.e. the deviation of gaseous rotation from stellar rotation, can be increased for longer time simply as a consequence of a large infalling gas cloud, as it may change the angular momentum of the warm gas. Nevertheless, bursts in AGN activity (not necessarily related to merger events, though) and associated AGN feedback, are necessary, to produce longer-term disordered gaseous motions and related permanent high levels of {\it all three} irregularity parameters as a consequence of heating and ejecting warm gas and to large extent preventing its re-accretion back onto the galaxy \citep[see][]{brennan2018}.

\begin{table*}[htb!]
\centering
\begin{tabular}{l | c c c c c c c}
     z & & $\mu$(Max. \kthree) & $\sigma$(Max. \kthree) & $\mu$(Max. \diffpa) & $\sigma$(Max. \diffpa) & $\mu$(\stdevpa) & $\sigma$(\stdevpa) \T \B \\
\hline
 1.3 & AGN    & 0.12 & 0.06 & 14.81 & 15.93 & 3.12 & 1.75 \T \B \\
  & NoAGN  & 0.11 & 0.05 & 20.08 & 19.07 & 3.76 & 2.87 \T \B \\
 \hline
 0.5 & AGN    & 0.30 & 0.42 & 52.70 & 56.81 & 7.66 & 7.79 \T \B \\
  & NoAGN  & 0.10 & 0.06 & 41.30 & 62.87 & 3.31 & 1.12 \T \B \\
\end{tabular}
\caption[Means and standard deviations of irregularity parameters for simulated galaxies]{\footnotesize Means and standard deviations of the maxima of irregularity parameters for simulated galaxies with and without AGN feedback at z = 1.3 and z = 0.5.}
\label{table:simssummary1}
\end{table*}

%\newpage
\subsection{Sample of 20 simulated massive galaxies}\label{simsstats}

In this section, we investigate the statistics of the irregularity parameters of our full galaxy sample, ran with and without AGN feedback. The irregularity parameters of each galaxy are calculated at 25 random projections. Since the maximum values of these parameters (computed within $< r_e$) are generally more sensitive to AGN activity, the warm gas content, and the orientation of each galaxy than the mean values, we will present the statistics of the maximum values for the remainder of this article.

Figs. \ref{fig:irrparamssims1} and \ref{fig:irrparamssims2} show \kthree\ vs. \stdevpa\ (top row), \diffpa\ vs. \stdevpa\ (middle row), and \diffpa\ vs. \kthree\ (bottom row) for all 10 simulated galaxies with (orange, left column) and without (blue, right column) AGN feedback, at z = 1.3 and z = 0.5, respectively. The irregularity parameters for edge-on orientation (i.e. uncontaminated by any spurious increase caused by face-on orientations, as explained in Section \ref{simscases}) are shown by the triangles. In order to additionally represent the range of irregularity parameters spanned for various galaxy orientations, the contours represent the kernel density estimates of the irregularity parameters for each galaxy in our sample at 25 different, randomly chosen orientations, color-coded by number density. 

A priori, we would expect galaxies with AGN feedback to have higher irregularity parameters than those without, as the gas in galaxies with AGN  is exposed to energy and momentum injection from the central accreting BH, in addition to other sources of disturbance such as mergers, stellar feedback etc. However, as we have also seen for the three test cases (Section \ref{simscases}), infalling gas may also lead to sustained high values of \diffpa, irrespective of AGN feedback.

Fig. \ref{fig:irrparamssims1} shows that at z = 1.3, AGN feedback is not (yet) strongly affecting the irregularity parameters: irrespective of feedback, their ranges covered by simulated galaxies at random projections are hardly exceeding the "regular" areas (dashed lines). Only  the \kthree\ parameter extends with AGN feedback up to slightly larger values (\kthree\ $<$ 0.5) than without AGN feedback (\kthree\ $<$ 0.2). Specifically for galaxies at edge-on projection (triangles), Fig. \ref{fig:irrparamssims1} additionally illustrates that the irregularity parameters of 7 and 6 out of 10 galaxies with and without AGN feedback, respectively, lie within the ``regular'' range (dashed lines) of the \kthree\ vs. \diffpa\ panels (top row). Fully irrespective of AGN feedback, 8 of 10 edge-on projected galaxies reside in the regular range of the \stdevpa\ vs \diffpa\  and of the  \stdevpa\ vs \kthree\ panels (middle and bottom rows). 

At z = 0.5, however, the situation changes, consistent with our a-priori expectation: the density contours of Fig. \ref{fig:irrparamssims2} illustrate that the ranges of all three irregularity parameters, in particular \stdevpa\ and \kthree, extent to significantly higher values for galaxies with AGN feedback than for those without this process, pointing towards a strong impact of AGN feedback on largely destroying any ordered rotation/motion of the warm gas content as well as distorting any alignment with the stellar motion. Specifically for galaxies in edge-on view, we find that with AGN feedback three, five and four out of ten galaxies reside within in the regular area of the \kthree\ vs. \diffpa,  the \stdevpa\ vs \diffpa\  and the  \stdevpa\ vs \kthree\ panels, respectively, whereas without AGN feedback the vast majority of galaxies have a regular kinematic features (7/10, 8/10 and 9/10, respectively).

Table \ref{table:simssummary1} summarises the means and standard deviations of the maximum values of the irregularity parameters for the edge-on projected galaxies at $z=1.3$ and $z=0.5$, pointing to a similar conclusion as discussed for Figs.  \ref{fig:irrparamssims1} and \ref{fig:irrparamssims2}:
The average irregularity parameters of galaxies at $z=1.3$ are similar with and without AGN feedback   (in fact, the kinematics of galaxies without AGN feedback can be even slightly less regular than those with AGN feedback), whereas at $z=0.5$ average irregularity parameters are higher with AGN feedback than without. This difference is more pronounced for \kthree\ and \stdevpa\ compared to \diffpa, as the latter parameter may also increase for galaxies without AGN feedback due to infalling gas clouds.

\begin{comment}
\begin{table*}
\centering
\begin{tabular}{l | c c c c}
   z  &  & \stdevpa\ vs. Max. \kthree\ & \stdevpa\ vs. Max. \diffpa\ & Max. \kthree\ vs. Max. \diffpa\ \T \B \\
\hline
 1.3 & AGN    & 0.746 & 0.613 & 0.194 \T \B \\
 & NoAGN  & 0.567 & 0.271 & -0.343 \T \B \\
 \hline
  0.5 & AGN    & 0.018 & 0.231 & -0.117 \T \B \\
 & NoAGN  & -0.063 & -0.073 & -0.196 \T \B \\
\end{tabular}
\caption[Correlation coefficients for combinations of irregularity parameters for simulated galaxies at z = 1.3 and z = 0.5]{\footnotesize The Pearson correlation coefficients for each pair of irregularity parameters for simulated galaxies with and without AGN feedback at z = 1.3 and z = 0.5.}
\label{table:simscorr1}
\end{table*}
\end{comment}

\section{Observational data from the CALIFA survey}\label{obs}

In this section, we briefly describe the source of our observed data (Sect. \ref{califa}), before detailing the means by which the data were cleaned and analyzed (Sect. \ref{obssample}).

\subsection{The CALIFA survey}\label{califa}

The Calar Alto Legacy Integral Field Area Survey (CALIFA, \citealt{sanchez2012,sanchez2016}) investigated a representative sample of galaxies in the Local Universe encompassing a wide range of morphologies and luminosities. 
It explored the diversity in basic characteristics such as the kinematics of their stars and gas, the emission properties of their ionized gas, and their stellar populations, spatially resolved on scales between 400 and 1200\,pc. 
The galaxies were diameter-selected to fit an extent of two half-light radii onto the spectrograph instrument field.
From an SDSS-based ``mother sample'' of 937 galaxies 667 were observed having redshifts between 0.005 and 0.03. 
Two thirds have late-type, disk-dominated morphologies, while we concentrate in our study on the $\sim$ 200 early-type galaxies (ETGs) to match the simulated objects that are all spheroidal at $z=0$. 

The observations were carried out using the PMAS/PPak spectrograph \citep{roth2005,kelz2006} mounted on the Calar Alto observatory's 3.5m-telescope. This instrument features a 74$\arcsec$ $\times$ 62$\arcsec$ hexagonal field of view
(FoV). 331 fibers of 2.7$\arcsec$ sample the FoV and 36 further fibers sample the sky near the FoV. Observations were carried out at low (V500, R $\sim$ 850 and spectral coverage between 3750 and 7500 \AA) and medium (V1200, R $\sim$ 1700 and spectral coverage between 3700 and 4200 \AA) spectral resolution, with each galaxy covered by two to three dithered exposures per mode. Once the raw data were taken, they were processed by the CALIFA pipeline \citep{husemann2013}, which removes cosmic rays, carries out flux calibration and extinction corrections, and interpolates the data onto a 78$\arcsec$ $\times$ 72$\arcsec$ grid. 
All processed data cubes were made public and provided to the community via regular Data Releases, the final one being the DR3 \citep{sanchez2016}.

\subsection{Observed data sample}\label{obssample}

For our study we selected 123 early-type CALIFA galaxies, whose V500 data were processed by {\sc Porto3D} in exactly the same way as described by \citet{papaderos2013} and \citet{gomes2016} 
for a subsample of 32 ETGs
that were examined with respect to the kinematics and emission characteristics of their warm gas component. 
{\sc Porto3D} is a pipeline designed to perform post-processing and spectral fitting of IFS data.
After data quality is assessed, individual spectra (per spaxel) are extracted, \textcolor{black}{deredshifted, rebinned to 1 \AA\ and fitted with the} publicly available stellar population synthesis code \textsf{STARLIGHT} (Cid Fernandes et al. 2005). Spectral fits were computed spaxel-by-spaxel using a base of simple stellar populations (SSPs) from both \citet{bruzual2003} and \textsf{MILES} \citep{sanchez-blazquez2006,vazdekis2010,falcon-barroso2011} covering a range of ages between 5 Myr and 13 Gyr. The results from these two modeling runs were combined in order to derive the most likely stellar population vectors and their main properties (e.g. stellar metallicities and ages). For each of these modeling runs, the best-fitting stellar continuum was then subtracted from each spaxel to create pure emission-line maps, which were in turn combined to improve on the determination of EWs and velocity centroids of emission lines.
{\sc Porto3D} incorporates a number of routines that allow for a precise subtraction of the  best-fitting stellar continuum, most importantly a continuum rectification technique that eliminates small-scale residuals in spectral fitting and permits reliable extraction of weak emission lines at a level of $<$ 1 \AA\ in equivalent width. 
This way two-dimensional maps with various quantities of interest were computed for all ETGs in our sample. These maps include the stellar and H$\alpha$ velocity and velocity dispersion, and their uncertainties, 
H$\alpha$ fluxes and EWs, and the \hbox{[N\,\textsc{ii}]}/H$\alpha$ flux ratio for every spaxel.  
The 
\com{interpolated spaxel scale of the CALIFA data (1$\arcsec$ $\times$ 1$\arcsec$)} corresponds on average to a spatial resolution of 0.5-1 kpc at the galaxies' typical distances (100-200 Mpc). As with the simulated galaxies, we do not perform any binning on the maps in order to preserve small-scale features.

We do, however, perform smoothing and masking on the velocity maps in the following manner. First, the H$\alpha$ velocity map (V$\alpha$) is median-filtered with a radius of 1 pixel (resulting in V$\alpha$\_med1) and 2 pixels (resulting in V$\alpha$\_med2), including the central pixel. Then, the relative absolute difference between the input velocity map and V$\alpha$\_med1 is calculated. Pixels with particularly high ($>$ 0.67, an empirically determined cutoff) relative absolute differences are set to 0 (bad), and pixels with very low relative absolute differences ($< 10^{-5}$) are also set to 0 because large, contiguous blocks of 
undefined velocities will all have the same value after median filtering, resulting in very low relative absolute differences. The remaining pixels are set to 1 (good), resulting in a ``mask'' differentiating between dubious and useful pixels. The resulting set of good pixels is then adaptively filtered with ESO-MIDAS using the algorithm by \citet{richter1991},
which smooths maps based on the local gradient. This smoothed image (V$\alpha$\_adap) is then also median-filtered (resulting in V$\alpha$\_adap\_med), and the absolute relative difference between V$\alpha$\_adap\_med and V$\alpha$\_adap is calculated. Finally, the pixels in the original map (V$\alpha$) with values of this second absolute relative difference greater than 0.67 are replaced by their corresponding values in V$\alpha$\_med2 (resulting in V$\alpha$\_pre\_smooth), and the bad pixels in V$\alpha$ are replaced with their respective values in V$\alpha$\_pre\_smooth. A final mask consisting of the previously mentioned 1 and 0 values (and further values, such as -1001 for pixels outside the field of view and $<$ -1001 for previously masked regions, such as foreground stars) is also output, and only the pixels marked 1 in the final, smoothed map are used for our analysis.

Of the 123 total galaxies (our \textit{full} sample), we further selected a sub-sample of 49 galaxies (the \textit{clean} 
sample) by visually inspecting both the original and smoothed H$\alpha$ velocity maps to ensure the presence of a relatively well defined, nearly contiguous  velocity field.
This was done in order to avoid a possible spurious increase in \kthree\ caused by gaps in the map (as explained above), but this may introduce a bias in favor of relatively gas-rich galaxies with an EW(H$\alpha$) of typically 1--3 \AA\ over their entire extent \citep[type~i and i+ ETGs in the notation by][and Gomes et al. 2016]{papaderos2013}.
In addition to disturbing the gas in a galaxy, AGN feedback can also remove it, and this effect is ignored if one excludes the gas-poorer galaxies, i.e. systems falling in the type~ii ETG class defined in \citet{papaderos2013}. 
Therefore, in the following, we will present statistics for both the full and the clean sample.
The smoothed H$\alpha$ velocity maps of these 49 galaxies are presented in the Appendix in Fig. \ref{fig:obsappendix}. The data were analyzed in the same manner as the data from the simulations (see Section \ref{sims}). We present the results of this analysis in the coming subsections. 

\begin{figure}[hb!] % whan
\includegraphics[width=0.999\columnwidth]{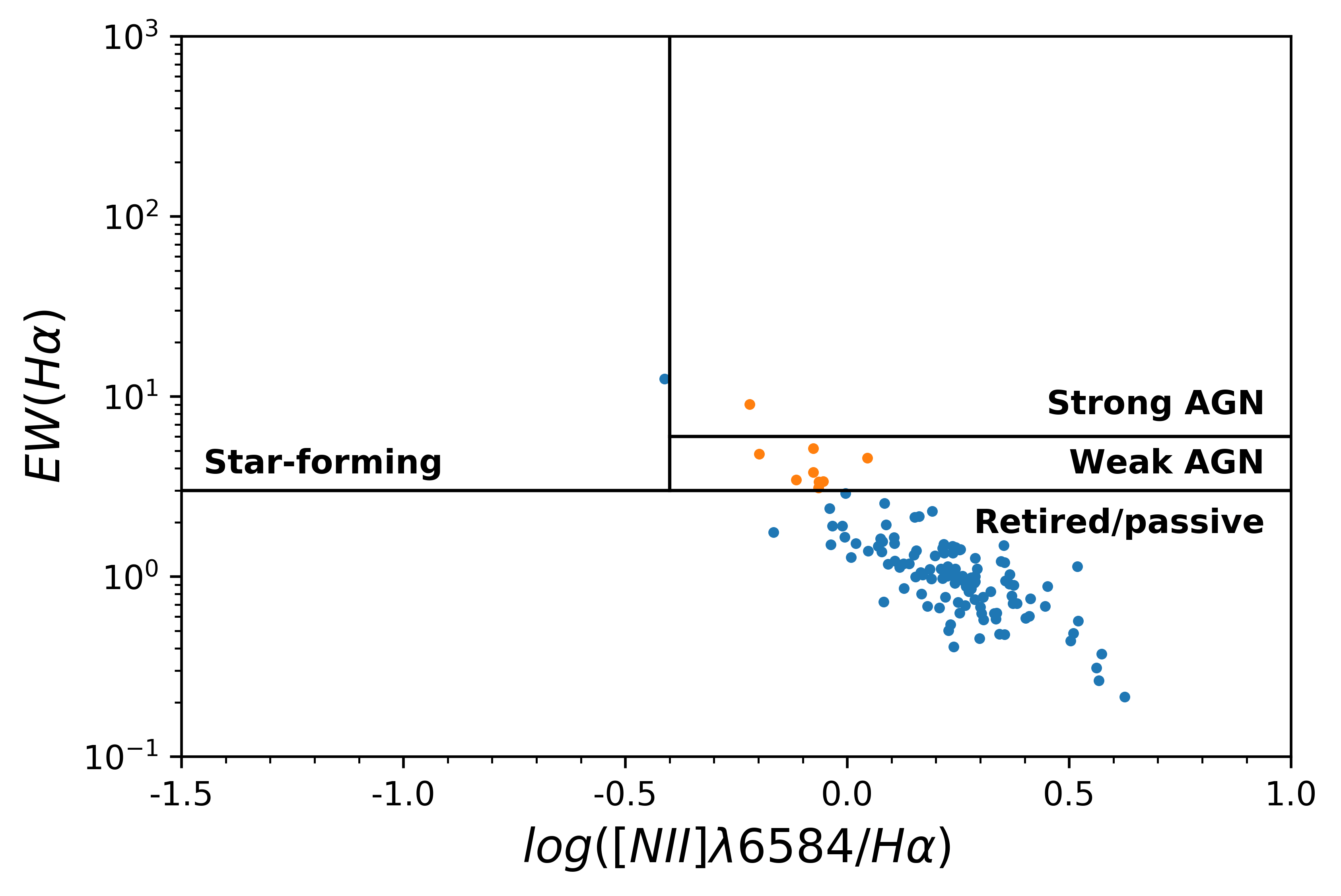}
\caption[WHAN diagram for observational data]{\footnotesize A WHAN diagram for our observational data sample. Galaxies characterized as either strong or weak AGN, and thus currently subject to AGN activity, are plotted in orange, and star-forming or passive/retired galaxies are plotted in blue. All nine galaxies currently classified as strong or weak AGN are part of the clean sample as well as the full sample.}
\label{fig:obswhan}
\end{figure}

We assume that AGN feedback is a ubiquitous process and thus that all the galaxies in our sample have been influenced by it at some point, but perhaps there is an enhanced effect if the galaxy is currently active. 
The activity status is assessed with a WHAN analysis (see Sect.~\ref{whantheory}),
allowing the division into strong and weak AGN, on the one hand, and the remaining inactive ones (retired and passive galaxies)
on the other (Fig. \ref{fig:obswhan}).
For this classification we use the mean nonzero \ewha\ and \n2ha\ values per spaxel for each galaxy. 
When we present the statistics of the values of the irregularity parameters in the coming subsections, we will present them separately for these two populations.

\section{Gas kinematics of CALIFA galaxies}\label{obsresults}

In this section, we begin by examining the values of irregularity parameters at specific radial bins for three test cases (Sect. \ref{obscases}) before investigating the statistics for the entire sample in Sect. \ref{obsstats}, as we did in Sect. \ref{simsstats}.

\subsection{Three case studies}\label{obscases}

\begin{figure*} % obs cases
\centering
\includegraphics[width=\linewidth]{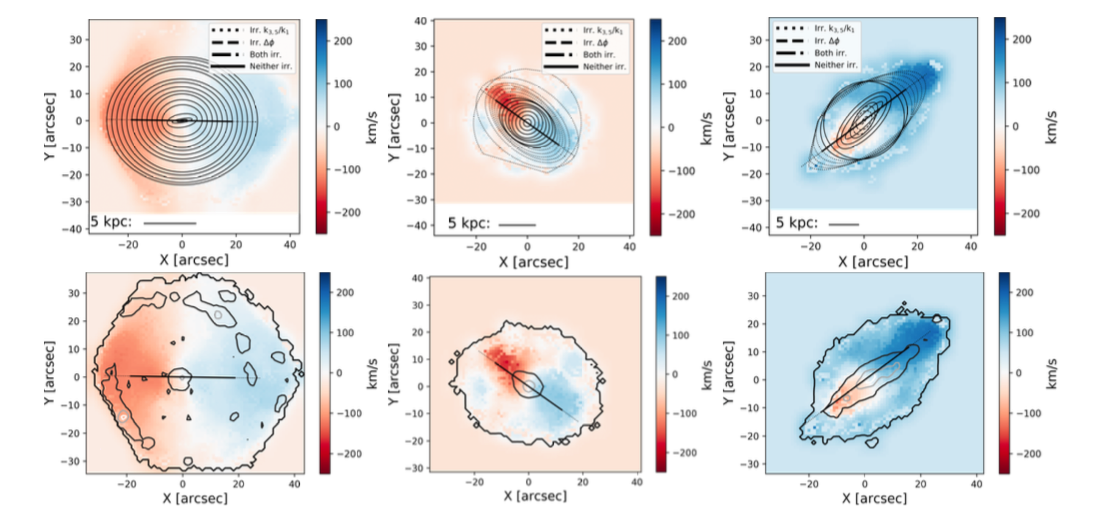}
\centering

\caption[Gas maps for NGC0932, NGC6146, and UGC10205]{\footnotesize For each of the observed galaxies NGC 0932 (left column), NGC 6146 (middle column), and UGC 10205 (right column), the velocity maps of the warm ionized gas without (top row) and with (bottom row) the contours of the H$\alpha$ flux overlaid (from darkest to lightest: the lowest nonzero value, 10\% of the maximum, 50\% of the maximum, 90\% of the maximum). The angle of the solid line in each map indicates the position angle of the median best-fitting ellipse across 1.5 r$_{e}$. The length of the solid line equals 1 r$_{e}$ and the length of the dotted line is equal to the remaining 0.5 r$_{e}$. The standard deviation of these ellipses' position angles is \stdevpa, and the position angles are compared with those of the stellar velocity to compute \diffpa. \kthree\ is calculated based on a different set of ellipses at the same radii with fixed position angles (all oriented along the solid line) and flattenings (the median flattening of the first set of ellipses). The linestyles of the ellipses correspond to the values of the irregularity parameters as follows: a solid line means that both \kthree\ and \diffpa\ are within the regular ranges, a dotted line means that \kthree\ is irregular, a dashed line means that \diffpa\ is irregular, and a dashed-dotted line means that both are irregular.}
\label{fig:obscases}
\end{figure*}

We present velocity maps of the warm gas of NGC 0932, NGC 6146, and UGC 10205 (Fig.\,\ref{fig:obscases}) 
and examine at which radial bin the irregularity parameters have values exceeding the threshold for undisturbed disk rotation.
The radial coverage between 0.1 r$_{e}$ and 1.5 r$_{e}$ is subdivided into 15 bins. 
In Table \ref{table:obssummarycases},
we list the means, maxima, and minima of the three irregularity parameters and the radii at which they show irregularity (if any).

\begin{table*}
\centering
\begin{tabular}{l c c c c c c}
     &  \kthree\ & Rad. Irr. & \diffpa\ & Rad. Irr. & \stdevpa\  \T \B \\
 \hline
 NGC 0932*  & 0.05$^{0.10}_{0.03}$ &  & 6.42$^{18.0}_{1.21}$ & & 4.26 \T \B \\
 \smallskip
 NGC 6146  & 0.28$^{1.17}_{0.03}$ & 9,10,11,12,13,14,15 & 20.43$^{40.36}_{2.84}$  & 11,12,13,14 & 10.95 \T \B \\
 UGC 10205* & 0.11$^{0.40}_{0.03}$ & 10,11,12,13,14 & 16.77$^{32.61}_{1.72}$  & 1,2,3 & 9.67 \T \B \\
\end{tabular}
\caption[Irregularity parameters for selected observed galaxies]{\footnotesize Mean, minimum and maximum values of irregularity parameters for selected observed galaxies as well as the radii at which the values are within the irregular range. Asterisks by galaxy names denote those currently affected by AGN activity, as determined by WHAN analysis.}
\label{table:obssummarycases}
\end{table*}

\subsubsection{NGC 0932 (Current AGN activity)}

NGC 0932 is, on average, regular in all three parameters at all radii. In fact, it is the only galaxy in our sample whose maximum values for all three parameters fall within the regular range. A visual inspection of the gas velocity map reveals a slight irregularity in the center, which may be due to the presence of an AGN (as indicated by WHAN analysis). The map of the H$\alpha$ flux reveals high values in the center of the galaxy as well as in a fragmentary ring-like structure around the galaxy's edge. This structure is located more than 1.5 r$_{e}$ away from the center and would thus not normally be analyzed. However, we extended the fitting to the galaxy's edge and found that \kthree\ was relatively high ($\sim$ 0.10) in the center, corresponding closely to the feature in the H$\alpha$ flux map, falling to $\sim$ 0.04 for most of the radial extent before rising to 0.09 again in the very outermost radius, apparently following the H$\alpha$ flux distribution. \diffpa\ also follows this trend, but not as closely. Overall, we find that \kthree\ and \diffpa\ react to disruption caused by AGN feedback in the center and stellar feedback in the outer ring in this galaxy.

\subsubsection{NGC 6146 (No current AGN activity)}

NGC 6146 is irregular with respect to \kthree\ at radii 9, 10, 11, 12, 13, 14, and 15 (corresponding to 13\farcs5, 15\farcs1, 16\farcs6, 18\farcs1, 19\farcs6, 21\farcs1, and 22\farcs6, respectively). It is irregular with respect to \diffpa\ at radii 11, 12, 13, and 14 (16\farcs6, 18\farcs1, 19\farcs6, and 21\farcs1), and its \stdevpa\ value is regular. 

Its H$\alpha$ flux map exhibits high flux in the center as well as in a faint cone-like feature extending toward the north-east. Here, \kthree\ and \diffpa\ follow the opposite of the trend from NGC 0932, in that their values are low in the center and rise toward the galaxy's edge. Perhaps here these two parameters are reacting more to the absence of warm gas than to disruption caused by stellar or AGN feedback. As we have seen, the parameters are sensitive to the amount of warm gas present as well as to sources of disruption such as feedback.

\subsubsection{UGC 10205 (Current AGN activity)}

The maximum values for \kthree\ and \diffpa\ fall outside the regular ranges  but in different regions for UGC 10205.
\kthree\ tends to be irregular in the middle to outer parts of the studied area (radii 10, 11, 12, 13, and 14, or 19\farcs4, 21\farcs3, 23\farcs3, 25\farcs2, and 27\farcs2) 
while \diffpa\ is irregular in the inner parts (radii 1, 2, and 3, or 1\farcs9, 3\farcs9, and 5\farcs8). 

There is a degree of twisting present in the inner portions of the gas kinematics, but not the stellar kinematics (which could indicate triaxiality with respect to kinematics in the same manner as isophotal twist), which would explain the increased values of \diffpa\ in the center. Indeed, there appears to be a lane of H$\alpha$-emitting gas present along the major axis of the galaxy, with a twist near the center (see Fig. \ref{fig:obscases}), which would explain these elevated \diffpa\ values. The elevated values of \kthree\ toward the edge of the galaxy may be a reaction to the decreased amounts of warm gas, as with NGC 6146.

\subsection{Observational sample of 49 CALIFA galaxies}\label{obsstats}

\begin{figure*} % irr params
\centering
\includegraphics[width=\linewidth]{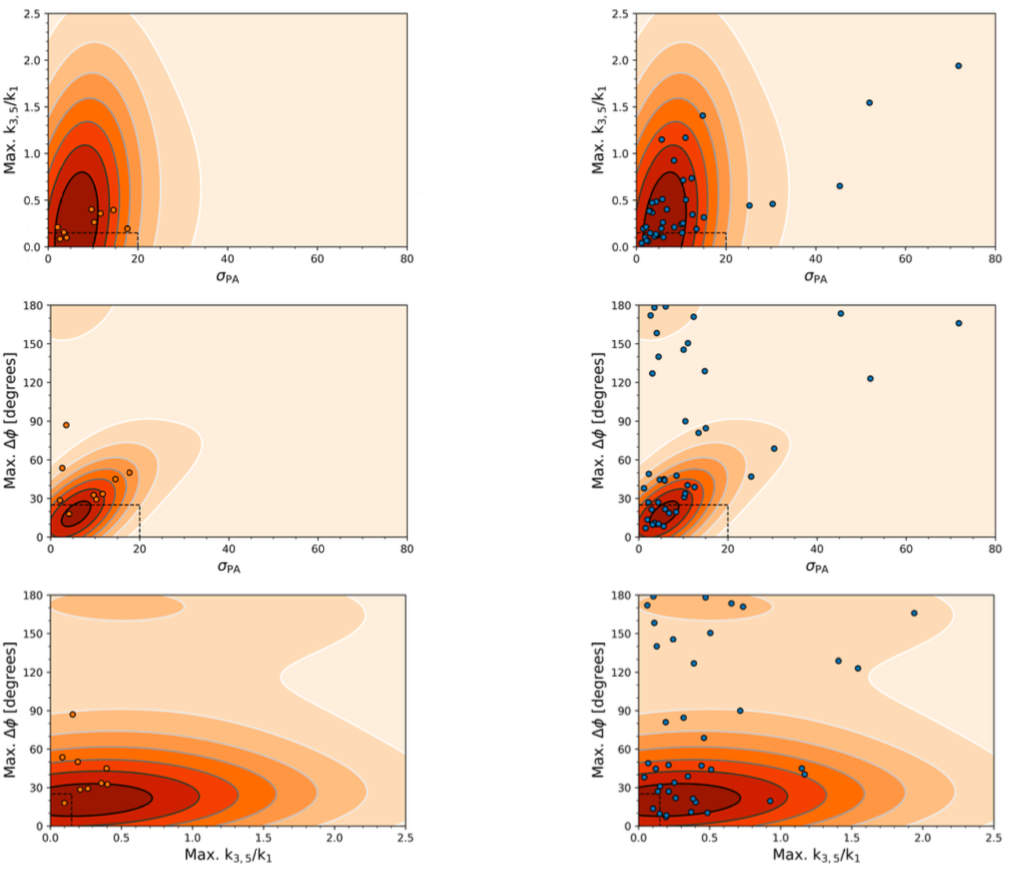}
\centering
\caption[Combinations of irregularity parameters for observed galaxies]{\footnotesize $\sigma_{\mathrm{PA}}$ vs. $k_{3,5}$/$k_1$ (top row), $\sigma_{\mathrm{PA}}$ vs. $\Delta \phi$ (middle row), and $k_{3,5}$/$k_1$ vs. $\Delta \phi$ (bottom row), for observed galaxies with (left column) and without (right column) current AGN activity. The colored contours represent the range of values exhibited by each simulated galaxy at 25 random orientations. Only galaxies simulated with AGN feedback were included. The round markers represent the maximum values within 1.5 r$_{e}$ for observed galaxies. The dashed lines indicate the upper boundaries of the parameter ranges within which a galaxy's kinematics are considered ``regular.''}
\label{fig:irrparamsobs2}
\end{figure*}

\begin{table*}
\centering
\begin{tabular}{l c c c c c c}
     &  $\mu$(\kthree) & $\sigma$(\kthree) & $\mu$(\diffpa) & $\sigma$(\diffpa) & $\mu$(\stdevpa) & $\sigma$(\stdevpa)  \T \B \\
\hline
  Clean sample, Current AGN  & 0.24 & 0.12 & 41.97 & 20.38 & 8.48 & 5.63 \T \B \\
 Clean sample, No Current AGN+RG    & 0.46 & 0.44 & 74.81 & 60.49 & 11.46 & 14.63 \T \B \\
 \hline
  Clean sample, Current AGN+RG  & 0.41 & 0.41 & 67.63 & 55.32 & 11.14 & 13.67 \T \B \\
 Clean sample, No Current AGN    & 0.49 & 0.61 & 95.79 & 107.83 & 5.52 & 4.05 \T \B \\
 \hline
 \hline
Full sample, Current AGN  & 0.24 & 0.12 & 41.97 & 20.38 & 8.48 & 5.63 \T \B \\
  Full sample, No Current AGN+RG  & 1.46 & 1.93 & 116.51 & 57.43 & 35.42 & 27.73 \T \B \\
 \hline
  Full sample, Current AGN+RG  & 1.33 & 1.93 & 108.94 & 59.58 & 31.97 & 27.40 \T \B \\
 Full sample, No Current AGN & 1.73 & 1.39 & 127.12 & 51.80 & 46.04 & 27.81 \T \B \\
\end{tabular}
\caption[Means and standard deviations of irregularity parameters for observed galaxies]{\footnotesize Means and standard deviations of the maxima of irregularity parameters for observed galaxies.}
\label{table:obssummary}
\end{table*}

In this section, we present the general statistics of our observed sample. Fig. \ref{fig:irrparamsobs2} shows an overview of the values of \kthree\ vs. \stdevpa\ (top row), \diffpa\ vs. \stdevpa\ (middle row), and \diffpa\ vs. \kthree\ (bottom row), for those galaxies in our clean sample which we determined via the WHAN analysis to be currently AGN-affected (left column) and not currently AGN-affected (right column). 
As already pointed out in Sect.~\ref{whantheory}, the latter (classified as retired or passive according to WHAN) fall in their majority on the locus of LINERs in BPT diagrams, and could therefore host some level of accretion-powered nuclear activity that due to extensive LyC photon escape evades detection through optical emission-line spectroscopy. 
The round markers represent the maximum values of the irregularity parameters within the observed radial range (1.5. r$_{e}$), and the dashed lines indicate the boundaries of the parameter ranges within which a galaxy’s kinematics are considered to be “regular” with respect to each parameter, in a similar manner to Figs. \ref{fig:irrparamssims1} and \ref{fig:irrparamssims2} in Sect. \ref{simsstats}. In order to facilitate comparison with the simulations, we have also included the kernel density estimate contours for the values of the irregularity parameters for the simulated galaxies with AGN feedback at 25 different, randomly chosen orientations at z = 0.5 from Fig. \ref{fig:irrparamssims2}. We performed the comparison in this manner because our simulated galaxies with AGN feedback at z = 0.5 are the closest proxies to our observed galaxies. This is because all observed galaxies are assumed to have felt the effects of AGN feedback, and z = 0.5 is the latest time at which all of the simulated galaxies in our sample have substantial amounts of warm gas. The comparison will be discussed in more detail in the following section.

Of the full sample of 123 galaxies, the harmonic fit performed by kinemetry did not converge for 9, and thus they are excluded from our analysis. Table \ref{table:obssummary} presents an overview of the mean values and standard deviations of the maximum values of the irregularity parameters within 1.5 r$_{e}$ for the full sample (\com{123} galaxies) and the clean sample (49 galaxies). In this table, we also present separate statistics for the case in which we include RGs in the sample bin with currently AGN-active and non-active galaxies. 

For the clean sample, current AGN activity does not enhance any irregularity parameters. In fact, it slightly reduces the observed parameter range. Of the clean-sample currently-AGN-affected galaxies, 1/9 is regular with respect to \kthree\ vs. \diffpa, 1/9 is regular with respect to \diffpa\ and \stdevpa, and 2/9 are regular with respect to \kthree\ and \stdevpa. The range of maximum \kthree\ values exhibited is 0.08-0.40, the range of maximum \diffpa\ values is 18.0$\degr$-87.08$\degr$, and the range of \stdevpa\ values is 2.11-17.68. Without current AGN activity, these numbers are 2/40, 10/40, and 10/40, respectively. The range of maximum \kthree\ values is 0.04-1.94, the range of maximum \diffpa\ values is 6.87$\degr$-179.03$\degr$, and the range of \stdevpa\ values is 1.15-71.82. The galaxies span the parameter ranges mostly evenly, with galaxies tending to cluster more closely at lower parameter values while still mostly remaining outside the regular ranges.

Both the means and the standard deviations of the maxima of the parameters are significantly lower for currently AGN-affected galaxies than for currently AGN-unaffected galaxies in the clean sample. We find variations of this result when we explore the full sample or include RGs %with currently AGN-affected or currently unaffected galaxies 
(lines 3-8 in Table \ref{table:obssummary}). In all cases, currently AGN-affected galaxies exhibit lower irregularity parameter values (and, in many cases, lower scatters) than currently non-AGN-affected galaxies.

A reason for this result could be that AGN feedback becomes more efficient after the peak in the accretion rate, as self-regulation means that feedback reduces the accretion rate. Thus, current AGN activity indicates both a lack of previous recent) feedback and a reservoir of gas available for accretion, which both tend to reduce the values of the parameters.

%\afterpage{\clearpage}

\begin{comment}
\begin{figure*} % irr params
\centering
%\text{Current AGN}\par
\includegraphics[width=\linewidth]{8.png}
\centering
\caption[Combinations of irregularity parameters for observed galaxies]{\footnotesize $\sigma_{\mathrm{PA}}$ vs. $k_{3,5}$/$k_1$ (top row), $\sigma_{\mathrm{PA}}$ vs. $\Delta \phi$ (middle row), and $k_{3,5}$/$k_1$ vs. $\Delta \phi$ (bottom row), for galaxies with (left column) and without (right column) current AGN activity. The colored contours represent the range of values exhibited by each simulated galaxy at 25 random orientations. Only galaxies simulated with AGN feedback were included. The round symbols represent the maximum values within 1.5 r$_{e}$ for observed galaxies.}
\label{fig:irrparamsobs1}
\end{figure*}
\end{comment}

\section{Comparison between simulations and observations and discussion of results}\label{comp}

In this section, we will compare the results from our observed and simulated data samples and discuss some caveats of our analysis.

Fig. \ref{fig:irrparamsobs2} shows that the observed galaxies follow the contours of the simulated galaxies with AGN feedback at different orientations fairly closely. 
As we have seen, it is important to keep in mind that orientations close to face-on can increase the values of the parameters (in both simulations and observations) as well. Nonetheless, we find that AGN feedback is a necessary process adopted in simulations to be more consistent with the range of observed irregularity parameters, whereas without AGN feedback the extent of irregularity parameters would be significantly too small.

In total, the observed galaxies tend to occupy wider parameter ranges. For the clean observed sample, the maximum values of \kthree\ ranged from 0.08 to 0.40 for currently AGN-affected and 0.04 to 1.94 for currently unaffected galaxies, or 0.04 to 1.94 overall, compared to 0.03 to 1.51 for the galaxies simulated with AGN feedback. The observed galaxies displayed \diffpa\ values of 18.0$\degr$ to 87.08$\degr$ (current AGN) and 6.87$\degr$ to 179.03$\degr$ (no current AGN), or 6.87$\degr$ to 179.03$\degr$ overall, compared to 4.81$\degr$ to 180.0$\degr$ for the simulated galaxies. Finally, the maximum values of \stdevpa\ ranged from 2.11 to 17.68 for currently AGN-affected and 1.15 to 71.82 for currently non-AGN affected clean-sample oberved galaxies, or 1.15 to 71.82 overall, compared to 1.26 to 28.01 for the simulated galaxies. The differences are most striking for \kthree, for which the parameter range is 28\% larger for observed galaxies, and \stdevpa, for which the parameter range is 164\% larger. We discuss possible reasons for these differences in the next subsection.

As we saw with the three observed case studies (Sect. \ref{obscases}), high irregularity parameter values correlate with regions of strong H$\alpha$ emission, such as central regions near an active nucleus or star-forming rings closer to the periphery, where stellar feedback distorts gas kinematics. High irregularity parameter values were also measured in regions with minimal flux in the outer areas of galaxies. As we saw with the simulated case studies in Sect. \ref{simscases}, the irregularity parameters (especially \kthree\ and \stdevpa) are sensitive to the level of energy input from feedback sources as well as the amount of warm gas present.

\subsection{Possible origin of different irregularity parameter ranges}\label{difforigin}

\textcolor{black}{
There are several reasons why we presume the observed galaxies present higher values 
of the irregularity parameters than the simulated galaxies.} 

%\textcolor{black}{
First, as several spatially resolved studies of ETGs with wide-FoV IFS arrays have revealed, these seemingly simple systems actually show a remarkable degree of morphological and kinematical complexity both in their WIM and stellar component 
\citep[e.g.,][]{gomes2016}. For instance, their stellar kinematics exhibit a wide heterogeneity -- ranging from  overall pressure-supported to rotational patterns, and even counter-rotating components or rotation along the galaxy's minor axis.
Likewise, their WIM also shows a kinematical diversity -- from a complete lack of coherent motions all the way to ordinary rotation throughout the galaxy extent, or within an oblate component \citep{sarzi2006}, with cases of kinematically distinct outflows or ionization cones 
\citep{kehrig2012},
in some cases referred to as geysers \citep{roy2018}.

\textcolor{black}{
Secondly, a quantification of the degree of irregularity in the WIM becomes even more challenging due to the variation of these characteristics across galactocentric radius, a fact that may witness a superposition of different gas excitation mechanisms, such as AGN-driven in/outflows and ionization cones with low-level star formation \citep[e.g.,][]{Trager2000,Salim2012}. 
For instance, \citet{papaderos2013} have described two main classes of radial \ewha\ profiles for ETGs: type~I systems ($\sim$40\% of their sample) show a radially constant \ewha\ of 0.5-2.4 \AA\ (hereafter \ewha$_{\rm pAGB}$), i.e. well within the range of predictions from pAGB photoionization models. To the contrary, type~ii ETGs ($\sim$60\%) are characterized by positive \ewha\ gradients with a central \ewha$<$0.5 \AA\ that gradually raises to \ewha$_{\rm pAGB}$ in the galaxy periphery. 
The faintness of line emission in the central parts of type~ii ETGs presumably reflects an inwardly decreasing (increasing) WIM density (LyC photon escape fraction $f_{\rm LyC}$), and obviously strongly impedes studies of gas kinematics.
The situation is further complicated by faint star-forming spiral-like features in the periphery of some type~i ETGs 
\citep[classified as i+ by][]{gomes2016} that locally give rise to \ewha\ values $>$10 \AA, i.e. several times larger than \ewha$_{\rm pAGB}$. Such fine features, traceable in i+ ETGs on scales of a few arcsec (e.g., \object{NGC 932}) are definitely beyond the resolution of our simulations.}

\textcolor{black}{
Furthermore, the inherent kinematical complexity of ETGs might be slightly enhanced by noise in the observational data.
This is because of the generally very faint emission (of the order of 0.5--3 \AA) in most of these systems that, depending on the quality of fitting and subtraction of the underlying stellar continuum, leaves uncertainties of typically no less than $\sim$20 km/s in velocity determinations for individual spaxels.
Even though such small-scale random uncertainties in $V_{\alpha}$ maps should leave irregularity parameters from \textsf{kinemetry} unaffected, an attempt was made to mimick a similar degree of noise into the simulated data by splitting each particle into 60 randomized pseudoparticles.
This has not, however, markedly altered the irregularity determinations for the simulated galaxy sample, as expected.}

Finally, it is also possible that the implementation of AGN feedback in the simulations needs to be refined. As stated previously, AGN feedback in simulations is necessary for suppressing late, in-situ star formation and thus obtaining realistic numbers of slow-rotating, massive early-type galaxies. 
However, the specific implementation of AGN feedback we used may still need to be reworked, in that the warm gas is currently being heated too strongly. The other sub-grid physics models that regulate, e.g., stellar feedback may also need to be adjusted.
Furthermore, observed galaxies are subject to additional mechanisms which can disturb their gas kinematics, such as winds driven by cosmic rays (e.g. \citealt{wiener2017}) or OB stars that migrate into low-density regions after their formation \citep{li2015}. These mechanisms were not included in the model used to generate our simulated galaxies, and their inclusion may have helped to bridge the gap.

Additionally, our spatial resolution in this study was set to 1\,kpc, which is the average physical scale of the observed CALIFA data, but
that could be insufficient for capturing part of the feedback phenomena expected from an AGN, as for example collimated gas outflows that might give rise to localized shocks and eventually trigger in situ star formation in the circumnuclear and peripheral zones of ETGs.
The dynamic range of simulations is currently limited, however. At the moment, it is impossible to simultaneously resolve the typical Jeans length of ionized gas and accurately model the potentially massive galaxies containing it. Stellar and AGN feedback as well as other processes, to the extent that they are understood, must be included in the form of empirically motivated sub-resolution implementations, omitting the cosmological context. 

Moreover, our choice of the times at which snapshots were extracted from the simulations (typically every 100-200 Myr) might have been too coarse
for optimally monitoring brief kinematical perturbations owing to AGN feedback.
Even though this time resolution is finer than the typical dynamical time of the WIM (several hundred Myr for most of the considered models), it might significantly exceed the timescale for the re-regularization of the warm gas kinematics via differential shearing and energy dissipation through cloud-cloud collisions, as well as the visibility timescale of AGN-driven outflows and associated shocks. 
Also, the timescale over which kinematical perturbations in the WIM are effectively detectable both in simulated and observed maps likely depends on the radial dependence of various physical properties (e.g., gas density and velocity dispersion) which are poorly known for ETGs.
As simulation and observation techniques improve, it should be possible to perform studies like those carried out in this work at ever higher resolution and, thus, to better understand the forces at work.

\section{Summary and conclusions}\label{conc}

In this article, we analyze the velocity fields of massive early-type galaxies for indications of
AGN feedback processes exploiting both a simulated and an observed data sample.
Zoom-in simulations with the  GADEGET3-based \textsf{SPHGal} code were investigated of 20 objects following
different merger and accretion histories conducted once with and once without full prescriptions
of AGN and their feedback processes.
The simulations with AGN feedback show in general realistic
gas properties throughout cosmic history, such as less ordered rotation at
z > 1 and a decreasing WIM mass fraction with cosmic epoch for AGN hosts.
Our 114 observed local galaxies from the \textsf{CALIFA} survey were subdivided into AGN active
and passive/retired galaxies according to a WHAN analysis.
A further subsample was created to contain only 49 high-quality velocity fields.

We examine the velocity fields (VFs) of the warm ionized gas on spatial scales of 1\,kpc 
with the \textsf{kinemetry} tool.
To quantify possible peculiar features in the VFs due to AGN feedback processes
we measure three irregularity parameters with certain thresholds above which distortions
are significant.
These parameters encompass 
\kthree: deviations from simple orderly rotation revealing distinct kinematic components, 
\diffpa: the mean angle between the orientations of the stellar and gas VFs, 
and \stdevpa: the standard deviation of the position angles of the gas VF at different radii, 
each measured over a radial extent spanning 1.5 r$_{e}$. 

From the simulations, the observed galaxies, and the comparison of both
we find that:

\begin{itemize}

    \item AGN feedback can significantly disturb regular warm gas kinematics and often increases the values of the irregularity parameters.
    \item Inclination can artificially increase the values of the irregularity parameters turning an
    object from edge-on towards face-on.
    \item There is no propagation of the maxima inside-out visible after a peak in the BHAR following
    subsequent snapshots. 
\textrm{
    However, the output intervals may have been chosen too small to see this effect.
    }
 
   \item Interaction, in particular major merger, and accretion processes can also corrupt the gas
    VF and, thus, increase the values of the irregularity parameters. 
    This increase generally vanishes quickly in the absence of AGN
    feedback as the gas disk settles into its new configuration. 
    \item This new configuration may be out of alignment with the stellar component, however,
    keeping \diffpa\  elevated. Therefore, \kthree\ and \stdevpa\ are more sensitive to the presence
    of AGN feedback than \diffpa. 
    %OBS
     \item The distribution of the WIM component in observed local galaxies can deviate from a
     smooth disk and be rather heterogeneous.
     \item Localized regions of strong H$\alpha$ emission (in central areas near an active nucleus or
     in star-forming rings further out) or particularly weak H$\alpha$ emission (in gas-poor areas
     near the edge of a galaxy) can also lead to elevated values of the irregularity parameters.
    \item Galaxies classified as currently AGN-affected or non-AGN-affected by means of WHAN
    analysis do not show statistically differences in the values of their irregularity parameters.
    A galaxy whose BH was active in the past may still have a disturbed VF in the present, even
    though the AGN's activity has ceased.
    %COMPARISON
    \item Simulations need a process like AGN feedback to be consistent with the large ranges of
    irregularity parameters spanned by observed massive CALIFA ETGs.
    \item Tensions between irregularity parameters of observed and simulated galaxies point
    towards deficiencies in modelling AGN feedback (or other baryonic processes).
\end{itemize}

This study has explored one way to measure the extent to which the warm gas kinematics of massive early-type galaxies can be disrupted by AGN feedback. 
However, since the vast majority of galaxies will feel the effects of BH activity at some point, and other phenomena such as mergers or stellar feedback can disrupt the gas VFs as well, the exact contribution of AGN feedback is hard to pinpoint. 
We reiterate that our sample of 20 simulated and 49 (clean sample) or 114 (full sample) observed galaxies provides only sparse statistics. 
Future investigations of the WIM kinematics should incorporate smaller spatial scales both in
observations and simulations. The latter will also benefit from shorter timesteps between snapshots.

\section*{Acknowledgements}

Based on observations collected at the Centro Astronómico Hispano Alemán (CAHA) at Calar Alto, operated jointly by the Max-Planck Institut für Astronomie and the Instituto de Astrofísica de Andalucía (CSIC).
This article is based on the master thesis of JF that can be requested in full from BZ.
We greatly acknowledge fruitful discussions with Thorsen Naab, Jeremiah P. Ostriker and Davor Krajnovi\'c.
\textrm{
We also thank the anonymous referee for suggestions that improved the readability of our manuscript a lot.
}
P.P. thanks the European taxpayer, who in the spirit of solidarity between EU countries is offering to Portugal financial resources 
for supporting, via the FCT (Funda\c{c}\~{a}o para a Ci\^{e}ncia e a Tecnologia) apparatus, a research infrastructure in astrophysics. 
He acknowledges European and Portuguese funding via FEDER through COMPETE by the grants UID/FIS/04434/2013 \& POCI-01-0145-FEDER-007672 and PTDC/FIS-AST/3214/2012 \& FCOMP-01-0124-FEDER-029170. Additionally, this work was supported by FCT/MCTES through national funds (PIDDAC) by grant UID/FIS/04434/2019. P.P. also acknowledges support by Investigador FCT contract IF/01220/2013/CP1191/CT0002 and by FCT/MCTES through national funds (PIDDAC) and by grant PTDC/FIS-AST/29245/2017.
J.M.G. is supported by the DL 57/2016/CP1364/CT0003 contract and acknowledges the previous support by the fellowships CIAAUP-04/2016-BPD in the context of the FCT project UID/-FIS/04434/2013 \& POCI-01-0145-FEDER-007672, and SFRH/BPD/66958/2009 funded by FCT and POPH/FSE (EC).

\bibliographystyle{aa} % style aa.bst
\bibliography{references}

%\appendix
\section[]{Appendix}
The following pages present an overview of the velocity maps of the simulated and observed galaxies in our sample. 

%\clearpage

%\begin{comment}
\begin{figure*} % irr params
\centering
\includegraphics[width=\linewidth]{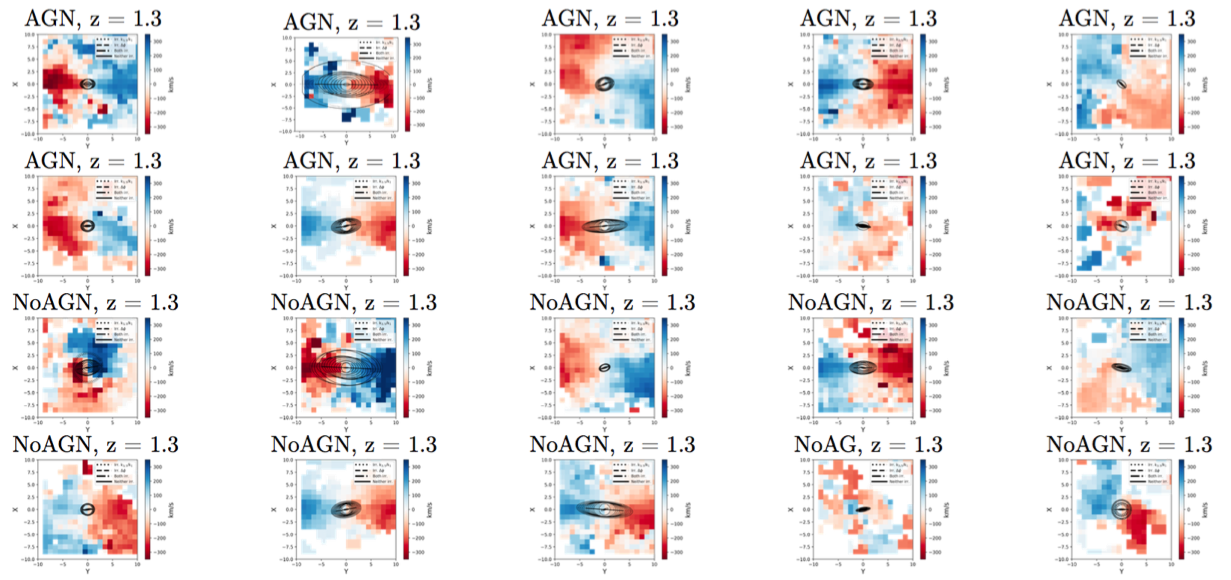}
\centering

\caption[Velocity maps of simulated galaxies at z = 1.3]{\scriptsize The velocity maps of the warm gas in the ten galaxies in our simulated data sample with (top two rows) and without (bottom two rows) AGN feedback at z = 1.3. Each galaxy is oriented edge-on, each map has an extent of 10 $\times$ 10 kpc, and the colors scale between -350 and 350 km/s. The line extending across the maps is oriented at the median position angle of the best-fitting ellipses within 1.5 r$_{e}$. The solid portion of the line has the length of the galaxy's effective radius from the center along the major axis, while the dotted portion represents the remaining fitted portion up to 1.5 r$_{e}$. The plotted ellipses are the best-fitting ellipses at every radius, with the ellipses' flattenings and position angles allowed to vary between radii. The standard deviation of these ellipses' position angles is \stdevpa, and the position angles are compared with those of the stellar velocity to compute \diffpa. \kthree\ is calculated based on a different set of ellipses at the same radii with fixed position angles (all oriented along the solid line) and flattenings (the median flattening of the first set of ellipses). The linestyles of the ellipses correspond to the values of the irregularity parameters as follows: a solid line means that both \kthree\ and \diffpa\ are within the regular ranges, a dotted line means that \kthree\ is irregular, a dashed line means that \diffpa\ is irregular, and a dashed-dotted line means that both are irregular.}
\label{fig:simsappendix1} 
\end{figure*}

\clearpage
\begin{figure*} % irr params
\centering
\includegraphics[width=\linewidth]{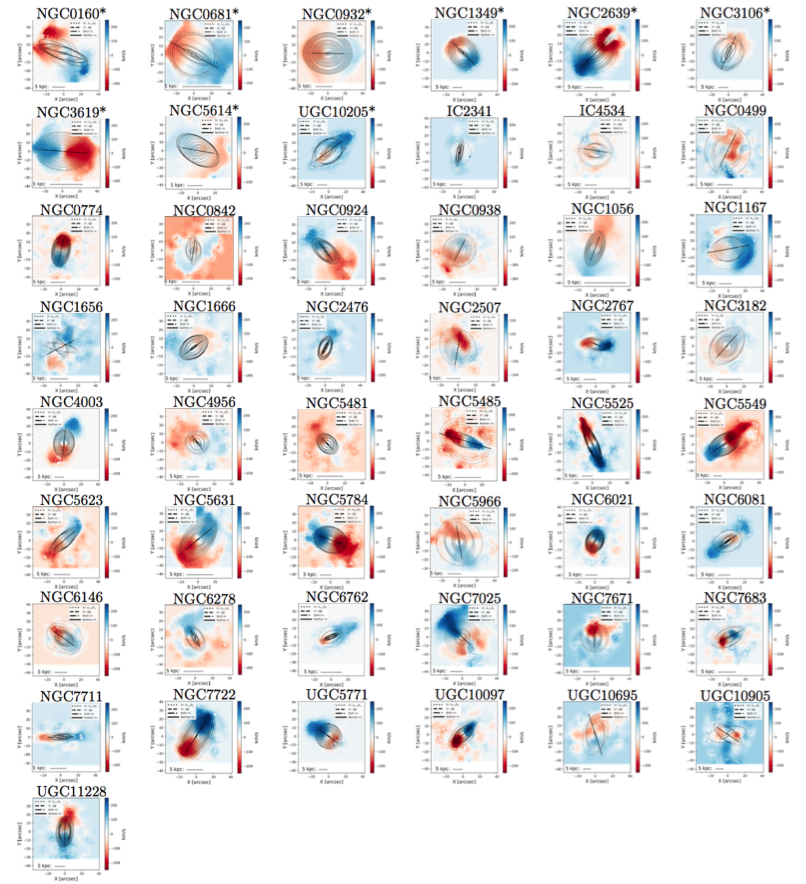}
\centering

\caption[Velocity maps of observed galaxies]{\scriptsize The velocity maps of the warm gas in the 49 galaxies in our observed data sample. Asterisks by galaxy names denote those currently affected by AGN activity, as determined by WHAN analysis. Only those data points contained in the ``masks'' (see text) are plotted. The star in each map marks the brightest continuum pixel, which was used as the center of the map for kinematic fitting. The line extending across the maps is oriented at the median position angle of the best-fitting ellipses within 1.5 r$_{e}$. The solid portion of the line has the length of the galaxy's effective radius from the center along the major axis, while the dotted portion represents the remaining fitted portion up to 1.5 r$_{e}$. The plotted ellipses are the best-fitting ellipses at every radius, with the ellipses' flattenings and position angles allowed to vary between radii. The standard deviation of these ellipses' position angles is \stdevpa, and the position angles are compared with those of the stellar velocity to compute \diffpa. \kthree\ is calculated based on a different set of ellipses at the same radii with fixed position angles (all oriented along the solid line) and flattenings (the median flattening of the first set of ellipses). The linestyles of the ellipses correspond to the values of the irregularity parameters as follows: a solid line means that both \kthree\ and \diffpa\ are within the regular ranges, a dotted line means that \kthree\ is irregular, a dashed line means that \diffpa\ is irregular, and a dashed-dotted line means that both are irregular.}
\label{fig:obsappendix} 
\end{figure*}
%\end{comment}

\end{document}